\documentclass[pdftex,twocolumn,epjc3]{svjour3}

\RequirePackage[T1]{fontenc}
\smartqed  

\RequirePackage{graphicx}
\RequirePackage{mathptmx}
\RequirePackage{flushend}
\RequirePackage[numbers,sort&compress]{natbib}
\RequirePackage[colorlinks,citecolor=blue,urlcolor=blue,linkcolor=blue]{hyperref}

\journalname{Eur. Phys. J. C}

\usepackage[below]{placeins}
\usepackage{flafter}
\usepackage{amsmath}
\usepackage{mathtools}
\usepackage{enumitem}
\usepackage{overpic}
\usepackage{tabularx}
\usepackage{upgreek}
\usepackage{dblfloatfix}
\usepackage{microtype}
\usepackage{notoccite}

\usepackage[defaultcolor=red]{changes}

\usepackage[switch]{lineno}

\usepackage{subfigure}
\usepackage{siunitx}
\usepackage{graphics}
\usepackage{booktabs}
\usepackage{orcidlink}

\begin{document}

\title{
Pulse shape simulation for the reduced charge collection layer in $p$-type high-purity germanium detectors
}

\author{
    P.~Zhang\thanksref{a1}\orcidlink{0009-0005-0472-0130}
    \and W.~Dai\thanksref{a1}\orcidlink{0000-0003-1732-7985}
    \and Q.~Zhang\thanksref{a1}\orcidlink{0009-0004-9929-640X}
    \and F.~Hagemann\thanksref{a2}\orcidlink{0000-0001-5021-3328}
    \and O.~Schulz\thanksref{a2}\orcidlink{0000-0002-4200-5905}
    \and C.~Alvarez-Garcia\thanksref{a2}\orcidlink{0000-0003-4768-8589}
    \and L.~Yang\thanksref{a1}\orcidlink{0000-0002-1859-3655}
    \and Q.~Yue\thanksref{a1}\orcidlink{0000-0002-6968-8953}
    \and Z.~Zeng\thanksref{a1}\orcidlink{0000-0003-1243-7675}
    \and J.~Cheng\thanksref{a1,a3}
    \and H.~Ma\thanksref{a1,cor}\orcidlink{0000-0001-8585-6665}
}
\thankstext{cor}{e-mail: mahao@tsinghua.edu.cn (corresponding author)}
\institute{
    Department of Engineering Physics, Tsinghua University, Beijing 100084, China \label{a1}
    \and
    Max-Planck-Institut für Physik, Garching, München 85748, Germany \label{a2}
    \and
    School of Physics and Astronomy, Beijing Normal University, Beijing 100875, China \label{a3}
}

\date{Received: date / Accepted: date}

\maketitle

\begin{abstract}
$P$-type high-purity germanium (HPGe) detectors are widely used across many scientific domains, and current data analysis methods have served well in many use cases. However, applications like low-background experiments that search for rare physics, such as dark matter, neutrinoless double-beta decay, and coherent elastic neutrino-nucleus scattering, could profit a lot from a more detailed understanding of the detector response close to the surface. The outer $n^+$ electrode of the $p$-type HPGe detector forms a layer with reduced charge collection, and events originating here can be a critical background source in such experiments. If the difference in detector pulse shape between detector surface and bulk events is known, it can be used to identify and veto these background events. However, a faithful simulation of the detector response in this surface region is difficult and has not been available as a standard method so far. We present a novel three-dimensional pulse shape simulation method for this reduced charge collection (RCC) layer. We have implemented this method as a new feature in the open-source simulation package \emph{SolidStateDetectors.jl} and show a validation of the numerical simulation results with analytical calculations. An experimental study using a $p$-type HPGe detector also validates our approach. The current implementation supports $p$-type HPGe detectors of fairly arbitrary geometry, but is easily adaptable to $n$-type detectors by adjusting the impurity density profile of the layer. It should also be adaptable to other semiconductor materials in a straightforward fashion.
\end{abstract}

\section{Introduction}\label{sec1}
$P$-type high-purity germanium (HPGe) detectors have been a mainstay in nuclear and particle physics for decades due to their excellent energy resolution. Applications that require a detailed understanding of the detector signal response still face challenges though: experiments like rare event searches can take advantage of the intrinsically low background of HPGe detectors. However, the intrinsic background of the detector and especially the surrounding parts is still not low enough for ultralow-background experiments such as direct dark matter detections~\cite{zhao_first_2013,jiang_limits_2018,ma_results_2020,li_limits_2013},
neutrinoless double-beta decay searches~\cite{dai_search_2022,zhang_searching_2024,abgrall_large_2017},
and coherent elastic neutrino-nucleus scattering detections~\cite{wong_research_2006,bonet_constraints_2021,yang_recode_2024,belov_new_2025}. Such experiments use detailed analysis of the detector signal shapes to distinguish signal from background events to further lower their effective background, but this is not an easy task.

The inhomogeneous electrical and weighting fields of large-volume HPGe detectors, together with the distinct drift behavior between bulk and surface regions, are both a blessing and a curse for such use cases: they result in different signal shapes for different event topologies that are associated with signal and background~\cite{martin_determining_2012,agostini_pulse_2022,comellato_topologies_2023,dai_virtual_2024}. However, they also make faithful simulation of such detectors a hard problem. Because of this, experiments have often relied on a data-driven approach to build digital pulse shape processing (DSP) procedures and statistical and machine-learning models for background discrimination~\cite{zhang_machine_2024,agostini_pulse_2022,holl_deep_2019,aalseth_maximum_2015,abt_alpha-event_2017,arnquist_alpha_2022,dai_virtual_2024,bonet_pulse_2024,radovic_machine_2018,cohen_machine_2018}. This data-driven approach has been successful, but is also limited, as the underlying truth of the data sets that the models are built on is not known event by event. Simulation-based and data-verified models would be the gold standard and have the potential for higher discriminatory power with better control of systematic effects.

Modern detector simulation packages like \emph{SolidStateDetectors.jl} (\emph{SSD.jl})~\cite{abt_simulation_2021,hagemann_determination_2024} can calculate detector fields and simulate charge drift in the detector bulk and can also be used to infer critical values like impurity profile parameters from measurements~\cite{abt_bayesian_2023}. A detailed simulation of HPGe near-surface effects on the charge drift, however, has not been easily available.

For $p$-type HPGe detectors in low-background experiments, the surface plays a special role.
The surface $n^+$ electrode of a $p$-type HPGe detector
is typically fabricated through thermal diffusion of lithium,
resulting in a heavily doped region with a thickness of \textasciitilde\,0.5--1\,mm.
However, this process introduces a layer
in the $n^+$ electrode that is almost inactive close to the surface, with its charge collection efficiency increasing as a function of depth: due to the weak electric field in this region and a higher trapping (or recombination) rate, charge carriers are prone to sudden termination, leading to incomplete charge collection~\cite{martin_determining_2012,aguayo_characteristics_2013,lehnert_background_2016,dai_modeling_2023}.
Thus, charge depositions in this layer are either not registered at all or exhibit a measured energy lower than their true energy. It is sometimes called dead layer, inactive layer, or transition layer in the literature. For clarity, we will call it the reduced charge collection (RCC) layer in this paper.
These RCC layer events constitute a significant background component in rare-event searches~\cite{aalseth_maximum_2015,bonet_pulse_2024,agostini_pulse_2022,li_differentiation_2014,yang_bulk_2018,li_identification_2022,zhang_machine_2024,wang_bulk-surface_2025}.

As the bulk material of the detector crystal is very radiopure, a large part of the background typically originates outside of the detector and on its surface. Background sources that emit alpha, beta, or low-energy gamma particles then deposit most of their energy in the detector close to the surface. As the layer has reduced charge collection, the energy of these events can be shifted into the signal region. Thus, it is important to distinguish them from actual signal events based on the shape of the signal pulse.

We want to study the charge drift of RCC layer events
and develop a pulse shape simulation method that can both deepen our understanding of the underlying detector physics and enable a more precise evaluation of the signal efficiencies
of pulse shape discrimination techniques~\cite{aalseth_maximum_2015,alvis_multi-site_2019,agostini_pulse_2022,comellato_topologies_2023,bonet_pulse_2024,dai_search_2022,zhang_searching_2024}. Realistic simulations can also generate clean datasets with known ground truth for developing advanced discrimination techniques~\cite{zhang_machine_2024}.

We chose the open-source modular software package
\emph{SolidStateDetectors.jl} \cite{abt_simulation_2021} as the technical basis for our RCC layer charge drift implementation. \emph{SSD.jl} is actively developed and maintained, has a modular and therefore extensible design, and already supports first-principle charge trapping and diffusion handling. It provided us with a solid basis to add mechanistic, non-heuristic RCC layer simulation. The code used to run the simulations in this paper is now freely available as of version 0.11.0 of
\href{https://github.com/JuliaPhysics/SolidStateDetectors.jl}{\emph{SSD.jl}}.
\section{Method}\label{sec2}
\begin{figure}[!htb]
    \centering
    \includegraphics[width=1\hsize]
    {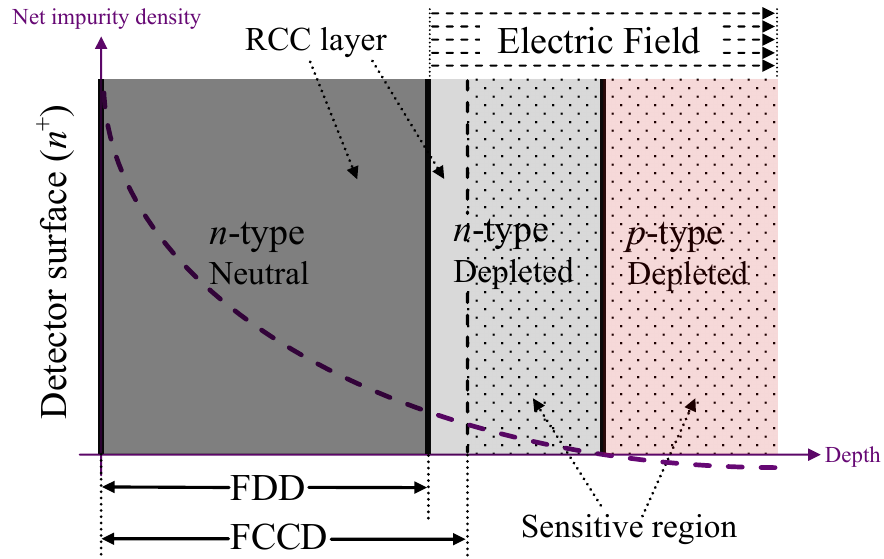}
    \caption{
    Schematic structure of the $n^+$ electrode of a $p$-type HPGe detector. 
    From the surface inward, there are $n$-type neutral region (dark gray),
    $n$-type depleted region (light gray),
    and $p$-type depleted region (pink).
    The net impurity density profile, illustrated as a purple dashed line, crosses zero at the $p$-$n$ boundary.
    The region with non-zero electric field is marked with parallel dashed lines.
    The sensitive region (full charge collection) is marked with dotted area,
    and the left near-surface region is the RCC layer.
    FDD and FCCD are also marked.
    }
    \label{fig.inactive_layer_structure}
\end{figure}

Figure \ref{fig.inactive_layer_structure}
illustrates the basic structure
of the $n^+$ electrode of a $p$-type HPGe detector.
Because of the thermal diffusion of lithium,
the donor concentration decreases rapidly from the surface inward,
reaching equilibrium with the acceptor concentration at the $p$-$n$ boundary.
When a reverse bias voltage is applied,
the detector is further depleted.
The depletion region extends outward from the $p$-$n$ boundary in both directions,
with the macroscopic electric field oriented toward the $p$-type region.
Consequently, the detector can be divided into three distinct regions
from the surface inward:

1) $n$-type non-depleted neutral region:
this region is characterized by an extremely high
concentration of donor impurities 
(\textasciitilde\,$10^{14}$--$10^{17}$\,cm$^{-3}$),
very short charge carrier lifetimes (O($\upmu$s)),
and no electric field~\cite{dai_modeling_2023}.
In this region, the donor impurities are ionized~\cite{huang_semiconductor_2012,knoll_radiation_2010},
but this region remains electrically neutral overall.

2) $n$-type depleted region:
here, charge carriers exhibit long lifetimes
(O(ms))~\cite{knoll_radiation_2010}.
The donor impurities are ionized and the carriers are depleted.
The boundary between the $n$-type depleted region
and the $n$-type neutral region is termed the full-depletion boundary,
where the electric field is exactly zero, see Figure \ref{fig.inactive_layer_structure}.
The electric field strength in this region increases
with the distance to the full-depletion boundary.
Following the convention of our previous research,
the depth from the detector surface to this boundary
is termed the full depletion depth (FDD)~\cite{dai_modeling_2023}.

3) $p$-type depleted region:
this region features a strong electric field
and long charge carrier lifetimes (O(ms))~\cite{knoll_radiation_2010}.

With "sensitive region", we refer to
the detector bulk volume with full charge collection efficiency (CCE).
RCC layer refers to the outside,
including the whole $n$-type neutral region
and a small part of the $n$-type depleted region, 
as shown in Figure \ref{fig.inactive_layer_structure}. 
Following the convention of previous researches~\cite{
aguayo_characteristics_2013,
gerda_collaboration_characterization_2019,
dai_modeling_2023},
the depth from the detector surface to the boundary
between the sensitive region and the RCC layer
is termed the full charge collection depth (FCCD),
which is also conventionally regarded
as the depth of the RCC layer.

The motion of charge carriers involves
four physical processes~\cite{hagemann_determination_2024,dai_pulse_2024,dai_modeling_2023}:

1) Drift under the electric field:
carriers undergo directional drift,
which only occurs in the depleted region with non-zero electric field.

2) Diffusion: random thermal motion of the carriers.

3) Self-repulsion: mutual repulsion between carriers
causes charge clouds to expand. This results in a correlation between total energy deposition and charge cloud size, which in turn can result in an energy dependence of the pulse shape.

4) Trapping (or recombination):
carrier drift may terminate suddenly
due to the trapping effect.

\subsection{Ionized impurity density profile}\label{sec2.1}
Ionized impurities refer to impurities with ionized atoms
regardless of whether the carriers are depleted, including donors and acceptors.
The ionized impurity density profiles govern the electric field and the physical processes within the RCC layer.

The initial acceptor impurity density is established during crystal growth, though external contaminants may enter during fabrication. This results in a complex three-dimensional distribution that is impossible to predict via first-principles modeling and difficult to measure experimentally after processing. Fortunately, this distribution can be inferred through capacitance-voltage measurements~\cite{hagemann_determination_2024}.
In contrast, donor impurities are introduced primarily through lithium thermal diffusion from the $n^+$ electrode.
The resulting profile is described by the complementary error function (erfc), representing the solution to the diffusion equation for a semi-infinite medium with a constant surface concentration.
The saturated surface lithium concentration and the diffusion coefficient
are modeled by empirical Arrhenius-type fits taken from the literature~\cite{sangster_ge-_1997,fuller_diffusion_1953,fuller_mobility_1954,dai_modeling_2023}.

Therefore, the net ionized impurity density profile throughout the crystal is described by
\begin{align}
&N_{\text{I,net}}(\hat{x}) = N_{\text{d}}(d)- N_{\text{a}}(\hat{x}) \,,  \label{eq.imp1}
\\[2mm]
&N_{\text{d}}(d) = N_{\text{s}} \cdot \text{erfc}(\frac{d}{2\sqrt{D_{\text{Li}}t_{\text{an}}}} ) \,,  \label{eq.imp2}
\\[2mm]
&N_\text{s} = N_{\text{s,0}} \cdot \exp(-\frac{T_0}{T_{\text{an}}} ) \,,  \label{eq.imp3}
\\[2mm]
&D_{\text{Li}} = D_{\text{Li,0}} \cdot \exp(- \frac{H}{RT_{\text{an}}} ) \,, \label{eq.imp4}
\end{align}
where $\hat{x}$ is the three-dimensional position,
$d$ is the depth from the $n^+$ surface,
$N_{\text{I,net}}(\hat{x})$ is the net ionized impurity density profile,
$N_{\text{d}}(d)$/$N_{\text{a}}(\hat{x})$ are the donor/acceptor impurity density profiles respectively,
$N_{\text{s}}$ is the saturated surface lithium concentration
described by an Arrhenius-type fit with the factors $N_{\text{s,0}} \approx 1.861\times10^{21}\,\text{cm}^{-3}$
and $T_0 \approx 6010\,\text{K}$~\cite{sangster_ge-_1997}, 
$t_{\text{an}}$ is the annealing time,
$T_{\text{an}}$ is the annealing temperature,
$D_{\text{Li}}$ is the lithium diffusion coefficient in germanium
described by another Arrhenius-type fit with the pre-exponential factor $D_{\text{Li,0}}$
and the activation energy $H$:
for $T_{\text{an}} \in [473\,\text{K}, 873\,\text{K}]$,
$D_{\text{Li,0}}$ = 2.5$\times$10$^{-3}$\,cm$^2$/s,
$H$ = \SI{11800}\,cal/mol~\cite{fuller_mobility_1954};
for $T_{\text{an}} \in [873\,\text{K}, 1273\,\text{K}]$,
$D_{\text{Li,0}}$ = 1.3$\times$10$^{-3}$\,cm$^2$/s,
$H$ = \SI{10700}\,cal/mol \cite{fuller_diffusion_1953}.
$R$ is the gas constant ($\approx$1.987\,cal/(K$\cdot$mol)).

\begin{figure}[!htb]
    \centering
    \includegraphics[width=1\hsize]
    {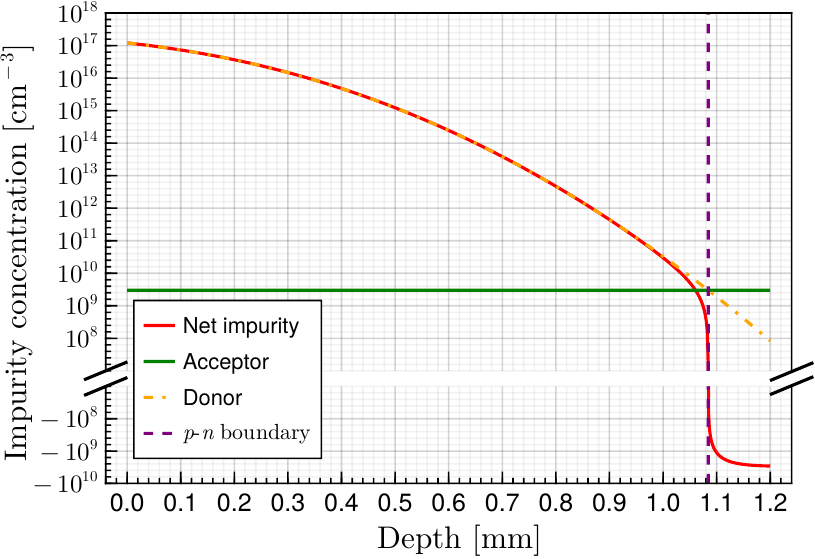}
    \caption{
    Donor (orange dashed), acceptor (green), and net (red) ionized impurity density profiles in the RCC layer 
    with $N_a$ = 3$\times$10$^9$\,cm$^{-3}$, $T_{\text{an}}$ = 623\,K, $t_{\text{an}}$ = 18\,min,
    and the saturated surface lithium concentration.
    The $p$-$n$ boundary is marked with a purple dashed line.
    }
    \label{fig.impurity_curves}
\end{figure}

Figure \ref{fig.impurity_curves} shows the impurity density profiles
in the RCC layer under typical parameter choices,
with the depth of the $p$-$n$ boundary
(where the net ionized impurity density equals zero)
marked by a purple dashed line.

\subsection{Charge-carrier mobilities}\label{sec2.2}
We consider three dominant factors that
affect the mobilities of carriers (electrons and holes) $\mu_{\text{e/h}}$ in our approach:
ionized impurity scattering ($\mu_{\text{e/h, I}}$),
neutral impurity scattering ($\mu_{\text{e/h, N}}$),
and acoustic phonon scattering ($\mu_{\text{e/h, A}}$)~\cite{mei_impact_2016, mei_impact_2017}.
Neutral impurities refer to unionized atoms,
such as carbon, oxygen, silicon, etc.,
which may originate from germanium crystal purification and growth processes~\cite{mei_impact_2016, haller_physics_1981, haller_carbon_1982}.
They are assumed to be uniformly distributed in the crystal.
The electrons/holes mobilities $\mu_{\text{e}}$/$\mu_{\text{h}}$ are calculated using
Equations \eqref{eq.emob1}--\eqref{eq.emob4} and 
Equations \eqref{eq.hmob1}--\eqref{eq.hmob4}, respectively~\cite{mei_impact_2016, mei_impact_2017, dai_modeling_2023}:
\begin{align}
&\frac{1}{\mu_{\text{e}}} = \frac{1}{\mu_{\text{e,I}}} + \frac{1}{\mu_{\text{e,N}}} + \frac{1}{\mu_{\text{e,A}}} \,, \label{eq.emob1}
\\[2mm]
&\mu_{\text{e,I}} = \frac{2.44 \times 10^{18} \cdot T^{1.5}/N_\text{I}}{\text{ln}(2.50 \times 10^{14} \cdot T^{2}/N_\text{I})} \,, \label{eq.emob2}
\\[2mm]
&\mu_{\text{e,N}} = \frac{3.00 \times 10^{19}}{N_\text{N}} \cdot (T^{0.5} + 1.93 \cdot T^{- 0.5}) \,, \label{eq.emob3}
\\[2mm]
&\mu_{\text{e,A}} = 9.32 \times 10^{7} \cdot T^{- 1.5} \,, \label{eq.emob4}
\\[2mm]
&\frac{1}{\mu_{\text{h}}} = \frac{1}{\mu_{\text{h,I}}} + \frac{1}{\mu_{\text{h,N}}} + \frac{1}{\mu_{\text{h,A}}} \,, \label{eq.hmob1}
\\[2mm]
&\mu_{\text{h,I}} = \frac{2.35 \times 10^{17} \cdot T^{1.5}/N_\text{I}}{\text{ln}(9.13 \times 10^{13} \cdot T^{2}/N_\text{I})} 
                   + \frac{1.51 \times 10^{18} \cdot T^{1.5}/N_\text{I}}
                   {\text{ln}(5.8 \times 10^{14} \cdot T^{2}/N_\text{I})} \,, \label{eq.hmob2}
\\[2mm]
&\mu_{\text{h,N}} = \frac{4.46 \times 10^{29}}{N_\text{N}} \cdot (T^{0.5} + 4.28 \cdot T^{- 0.5}) \,, \label{eq.hmob3}
\\[2mm]
&\mu_{\text{h,A}} = 7.77 \times 10^{7} \cdot T^{- 1.5} \,, \label{eq.hmob4}
\end{align}
where
$\mu_{\text{e}}$, $\mu_{\text{e,I}}$, $\mu_{\text{e,N}}$, $\mu_{\text{e,A}}$,
$\mu_{\text{h}}$, $\mu_{\text{h,I}}$, $\mu_{\text{h,N}}$, and $\mu_{\text{h,A}}$
have units of cm$^{2}\cdot$V$^{-1}\cdot$s$^{-1}$,
$N_\text{I}$ ( $ = N_\text{a} +  N_\text{d}$) is the ionized impurity density [cm$^{-3}$],
$N_\text{N}$ is the neutral impurity density [cm$^{-3}$],
and $T$ is the crystal temperature [K].

Concentration of neutral impurities is determined by matching 
the calculated sensitive-region hole mobility (using above equations)
to the experimentally measured value 
\cite{mei_impact_2016, mei_impact_2017, dai_modeling_2023}.
Under typical parameters
(sensitive-region hole mobility adopts the IEEE standard --- \SI{42000}\,cm$^{2}\cdot$V$^{-1}\cdot$s$^{-1}$
and $T= 90\,\text{K}$),
the neutral impurity density is determined to be 5.677$\times10^{15}$\,cm$^{-3}$.

\begin{figure}[!htb]
    \centering
    \includegraphics[width=1\hsize]
    {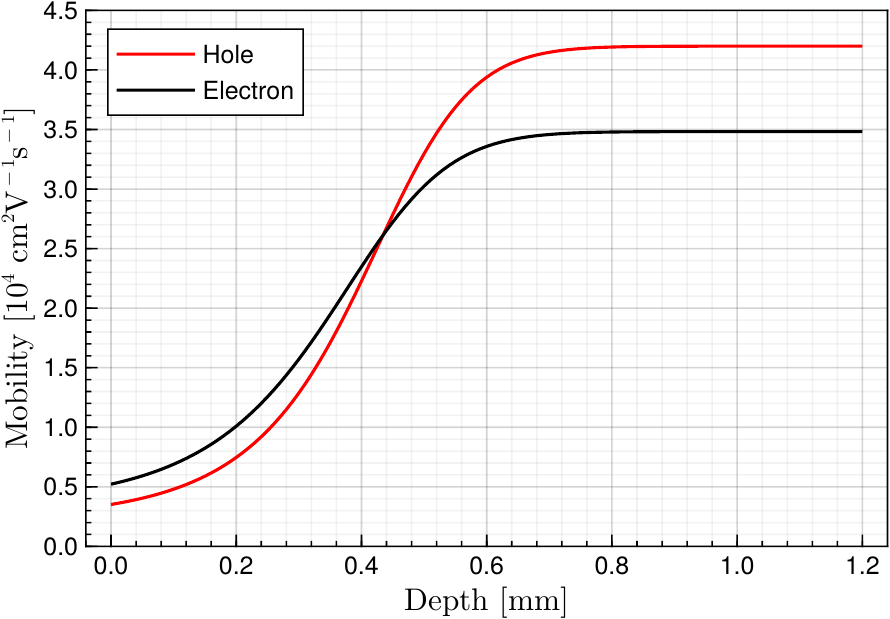}
    \caption{
        Mobilities of electrons (black) and holes (red) in the RCC layer
        when adopting the IEEE standard value of hole mobility in the sensitive region
        (\SI{42000}\,cm$^{2}\cdot$V$^{-1}\cdot$s$^{-1}$)
        and $T = 90$\,K.
    }
    \label{fig.mobility_curves}
\end{figure}

Figure \ref{fig.mobility_curves} presents the electron and hole mobility curves calculated using the parameters described above.
When getting closer to the $n^+$ surface,
the heavily doped lithium impurities significantly reduce
the carrier mobility compared to the bulk sensitive region due to impurity scattering.

It is important to note that this work treats the mobility as a low-field quantity and neglects velocity saturation.
In high-field regions, such as the area near the groove where the field can reach several kV/cm, this approximation leads to an overestimation of the drift velocity. However, because the total drift time for RCC-layer events is dominated by the slow transport within the heavily doped RCC layer, omitting the field dependence should introduce only a tiny bias to the overall signal shape. A more comprehensive drift model that incorporates both field-dependent mobility and crystal orientation effects is reserved for future work.

\subsection{Charge-carrier diffusion, self-repulsion, and trapping}\label{sec2.3}
The charge carrier diffusion process is simulated as a random walk
with a step length depending on the diffusion coefficients~\cite{hagemann_determination_2024}.
The diffusion coefficients $D_\text{e/h}$
are calculated from the mobilities $\mu_{\text{e/h}}$
using the Einstein relation~\cite{hagemann_determination_2024,dai_pulse_2024,dai_modeling_2023}:
\begin{equation}
D_\text{e/h} = \frac{k_\text{B}T}{e} \cdot \mu_\text{e/h} \,,
\label{eq.einstein}
\end{equation}
where $k_\text{B}$ is the Boltzmann constant and $e$ is the elementary charge.

The carrier self-repulsion effect is simulated by computing the electric
field contribution of other carriers and superimposing them onto the
static electric field at each drift step for each carrier~\cite{hagemann_determination_2024}.
The total electric field experienced by the $k$-th carrier at time $t$ is
given by
\begin{equation}
{\widehat{E}}_{k}({\widehat{r}}_{k}(t)) = \widehat{E}({\widehat{r}}_{k}(t)) +
    \sum_{j \neq k}^{}\frac{q_{j}}{4\pi\epsilon(\widehat{r}_{k} ) \left|\widehat{r}_{k} - \widehat{r}_j \right|^2} \frac{\widehat{r}_k - \widehat{r}_j}{\left|\widehat{r}_k - \widehat{r}_j \right|}
\,,
\label{eq.self_repulsion}
\end{equation}
where ${\widehat{r}}_{k}(t)$ represents the position of the
$k$-th carrier at time $t$,
$\widehat{E}({\widehat{r}}_{k}(t))$ denotes the static
electric field at that position, $q_{j}$ is the charge of the
$j$-th carrier, and $\epsilon(\widehat{r}_k )$ is the permittivity.

The trapping effect of carriers can be described by a lifetime model.
Ideally, the carrier lifetime should be treated as depth-dependent,
varying with the density of trapping centers.
In this work, as a first-order approximation,
we employ a piecewise constant-lifetime trapping model~\cite{dai_modeling_2023,boggs_numerical_2023},
assume different effective lifetimes in the sensitive region and the RCC layer.
Therefore, our trapping model utilizes four parameters:
electron/hole lifetimes in the RCC layer $\tau_\text{e/h,R}$,
and in the sensitive region $\tau_\text{e/h,S}$, respectively.
While the simplified lifetime profile is a necessary approximation that has also proven effective in Section \ref{sec4.4},
it introduces a potential uncertaintiy that is currently challenging to assess.
This limitation is discussed further in Section \ref{sec4.4}.
In addition, to mitigate the high computational cost of simulating trapping effects via pure Monte Carlo sampling, this work adopts an analytical approach in which an exponential decay is applied after the charge drift.

For this paper, we set $\tau_\text{e/h,S}$ to 1\,ms~\cite{knoll_radiation_2010},
a typical value for high-purity germanium.
However, in the RCC layer,
the extremely high donor impurity density makes
carriers more prone to trapping,
resulting in significantly shorter lifetimes
than those in the sensitive region.
The lifetimes $\tau_\text{e/h,R}$ must be determined by experimental calibration.
According to the Shockley-Ramo theorem~\cite{he_review_2001},
the induced signal $Q_{\text{e/h}}(t_i)$ is given by
\begin{align}
Q_{\text{e/h}}(t_i) &= Q_{\text{e/h, A}}(t_i) + Q_{\text{e/h, TR}}(t_i) + Q_{\text{e/h, TS}}(t_i) \,, \\[2mm]
Q_{\text{e/h, A}}(t_i) &= w(\widehat{r}_{\text{e/h}}(t_i))q_{\text{e/h}}(t_i) \,, \\[2mm]
Q_{\text{e/h, TR}}(t_i) &= \sum_{i^{\prime} \in I_\text{R}(t_{i-1})}^{}{w(\widehat{r}_{\text{e/h}}(t_{i^{\prime}}))q_{\text{e/h}}(t_{i^{\prime}})\frac{dt_{i^{\prime}}}{\tau_\text{e/h, R}}} \,, \\[2mm]
Q_{\text{e/h, TS}}(t_i) &= \sum_{i^{\prime} \in I_\text{S}(t_{i-1})}^{}{w(\widehat{r}_{\text{e/h}}(t_{i^{\prime}}))q_{\text{e/h}}(t_{i^{\prime}})\frac{dt_{i^{\prime}}}{\tau_\text{e/h, S}}} \,, \\[2mm]
q_{\text{e/h}}(t_i) &= q_{\text{e/h}}(t_0) - q_{\text{e/h, TR}}(t_i) - q_{\text{e/h, TS}}(t_i) \,, \\[2mm]
q_{\text{e/h, TR}}(t_i) &= \sum_{i^{\prime} \in I_\text{R}(t_{i-1})}^{}q_{\text{e/h}}(t_{i^{\prime}})\frac{dt_{i^{\prime}}}{\tau_\text{e/h, R}} \,, \\[2mm]
q_{\text{e/h, TS}}(t_i) &= \sum_{i^{\prime} \in I_\text{S}(t_{i-1})}^{}q_{\text{e/h}}(t_{i^{\prime}})\frac{dt_{i^{\prime}}}{\tau_\text{e/h, S}} \,,
\end{align}
where $t_i$ is the $i$-th timestamp of a drift (starting at $t_0=0$), 
$dt_i$ ($=t_{i+1}-t_i$) is the $i$-th time step,
$Q_{\text{e/h, A}}(t_i)$ denotes the induced signal contributed by alive carriers (at the timestamp $t_i$, the same below),
$Q_{\text{e/h, TR}}(t_i)$ and $Q_{\text{e/h, TS}}(t_i)$ denote the induced signal contributed by carriers trapped in the RCC layer and sensitive region respectively,
$q_{\text{e/h}}(t_i)$ denotes the charges of alive carriers,
$q_{\text{e/h, TR}}(t_i)$ and $q_{\text{e/h, TS}}(t_i)$ denote the charges of carriers trapped in the RCC layer and sensitive region respectively,
$\widehat{r}_{\text{e/h}}(t_i)$ denotes the carrier drift paths,
$w(\widehat{r}_{\text{e/h}}(t_i))$ denotes the weighting potentials,
and $I_\text{R}(t_i)$ and $I_\text{S}(t_i)$ denote the sets of timestamp indices when the carrier is located in the RCC layer and the sensitive region respectively.

\subsection{Charge collection efficiency}\label{sec2.4}
Based on the aforementioned physical models,
we can simulate both the paths of carriers
and the induced signals of RCC layer events.
Furthermore, we can calculate the CCE curve of
the RCC layer $f_\text{CCE}(x)$ and its uncertainty~\cite{dai_modeling_2023,ma_study_2017} with
\begin{align}
f_\text{CCE}(x) &= \frac{A_{\text{p}}(x)}{Ne} = \frac{A_{\text{p}}(x)}{\frac{E}{E_0}e} \,,
\label{eq.cce_calc}
\\[2mm]
\frac{\sigma_{f}}{f} &= \frac{1}{\sqrt{fN}}  \,,
\label{eq.cce_err_calc}
\end{align}
where $x$ is the depth of the simulated single-site event from the detector surface,
$A_{\text{p}}$ is the amplitude of the simulated charge pulse,
$E$ is the energy deposited,
$N$ is the initial number of the simulated charge carrier pairs,
and $E_0$ is the ionization energy for germanium. 
It should be noted that the uncertainty modeled here is overestimated, as the trapping effect is treated analytically after the drift.

The CCE curve is a fundamental property of $p$-type germanium detectors, mainly governed by the impurity density profiles and detector geometry. The applied bias voltage has only a minor impact on the CCE near the depletion boundary.
It has been proven to be crucial for interpreting experimental spectra and correcting distortions in the raw simulated energy spectra in the low-energy region
because alpha, beta, and low-energy gamma particles from outside have a higher probability to deposit energy close to the surface~\cite{
ma_study_2017,
dai_modeling_2023,
aalseth_maximum_2015,
agostini_pulse_2022,
bonet_pulse_2024
}.

\begin{figure}[!htb]
    \centering
    \includegraphics[width=0.9\hsize]{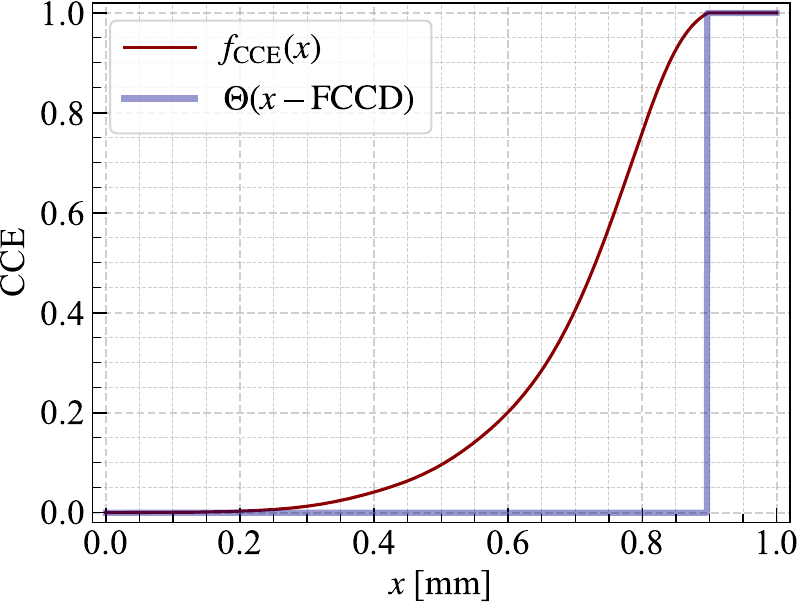}
    \caption{
        Schematic illustration of the CCE curve $f_\text{CCE}(x)$
        and the step function $\Theta(x-\text{FCCD})$ with FCCD = 0.9\,mm.
    }
    \label{fig.cce_step}
\end{figure}

For each event, we denote the simulated deposited energies and depths with $(E_i, x_i)$.
The raw simulated total energy $E_{\text{raw}}$,
the total energy corrected with CCE curve $E_{\text{w/}}$,
and total energy corrected without CCE curve $E_{\text{w/o}}$
are given by
\begin{align}
    E_{\text{raw}} &= \sum_i{E_i} \,,\\[2mm]
    E_{\text{w/}} &= \sum_i{f_\text{CCE}(x_i)E_i} \,,\\[2mm]
    E_{\text{w/o}} &= \sum_i{\Theta(x_i-\text{FCCD})E_i} \,,
\label{eq.cce_correction}
\end{align}
where $\Theta(x-\text{FCCD})$ is the step function
with discontinuity at $x$ = FCCD.
The $f_\text{CCE}(x)$ and $\Theta(x-\text{FCCD})$ are illustrated
in Figure \ref{fig.cce_step}.

For $p$-type HPGe detectors, in most cases,
electrons in the RCC layer contribute negligibly to
the induced signal on the $p^+$ contact
because the weighting potential there is nearly zero~\cite{li_identification_2022,dai_pulse_2024,zhang_searching_2023}.
Thus, the CCE approximately equals the fraction of holes that
escape the RCC layer before getting trapped or recombined,
reach the sensitive region, and get collected by the $p^+$ electrode.
One can determine the FCCD through pulse shape simulation but also
experimentally, by matching the experimentally measured energy spectrum
with the energy spectrum simulated using a Monte-Carlo software like Geant4~\cite{jiang_measurement_2016,ma_study_2017,aguayo_characteristics_2013,gerda_collaboration_characterization_2019,agostinelli_geant4simulation_2003,allison_geant4_2006,allison_recent_2016}.
\section{Method validation}\label{sec3}
We validate our simulation method by comparing the simulated CCE curve with a theory-based analytical calculation~\cite{dai_modeling_2023}.

In the theory-based approach, we compute the CCE in the RCC layer by
solving the one-dimensional hole transport equation analytically.
For this, we consider three hole-transport physics processes:
drift under the electric field, diffusion, and trapping.

In the numerical simulation, we disable the self-repulsion process and the signal contribution from electrons here, for direct comparison with the analytical results.

\subsection{Test geometry}\label{sec3.1}
\begin{figure}[!htb]
    \centering
    \includegraphics[width=1\hsize]{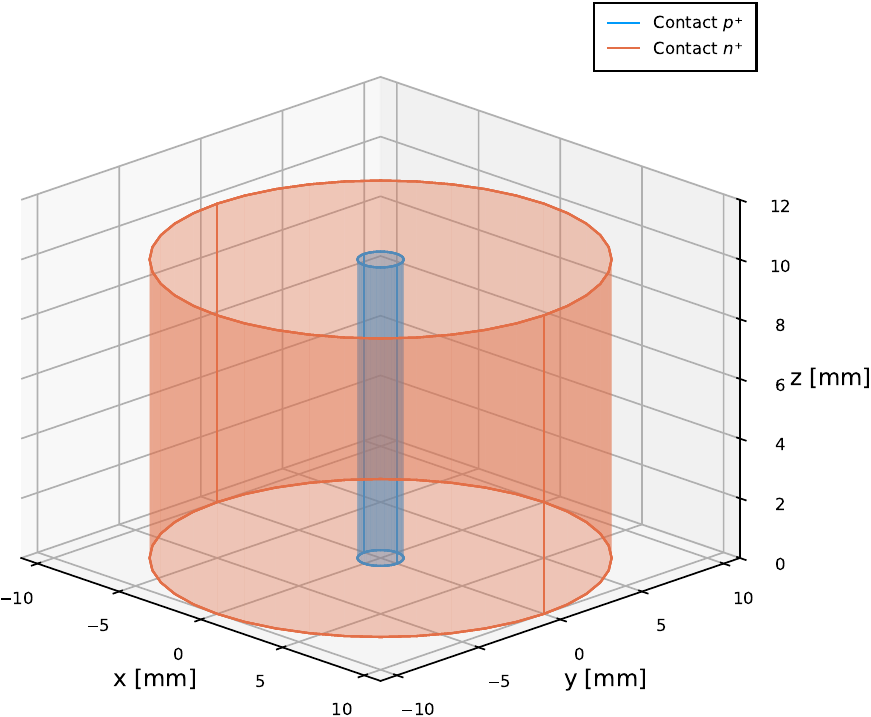}
    \caption{
        Visualization of the simulated hypothetical
        true-coaxial HPGe detector geometry for the
        analytical validation.
    }
    \label{fig.truecoaxial_det}
\end{figure}

We perform our validation
on a hypothetical true-coaxial HPGe
detector with a 1\,mm inner radius, 10\,mm outer radius, and 10\,mm height,
with the coordinate origin set at the center of the bottom surface of the detector,
see Figure \ref{fig.truecoaxial_det}. While not a common detector geometry in practice, the high symmetry enables an analytical solution of the problem.

\begin{figure*}[!htb]
    \centering
    \subfigure[]{
        \includegraphics[height=4.4cm]{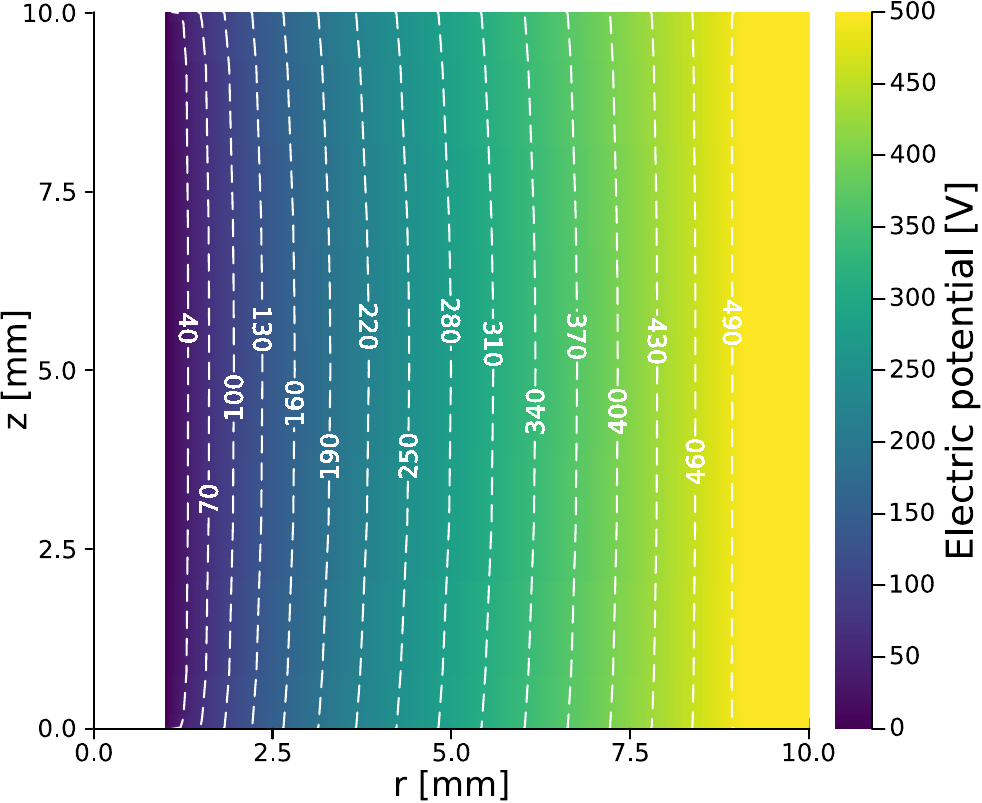}
    }
    \subfigure[]{
        \includegraphics[height=4.4cm]{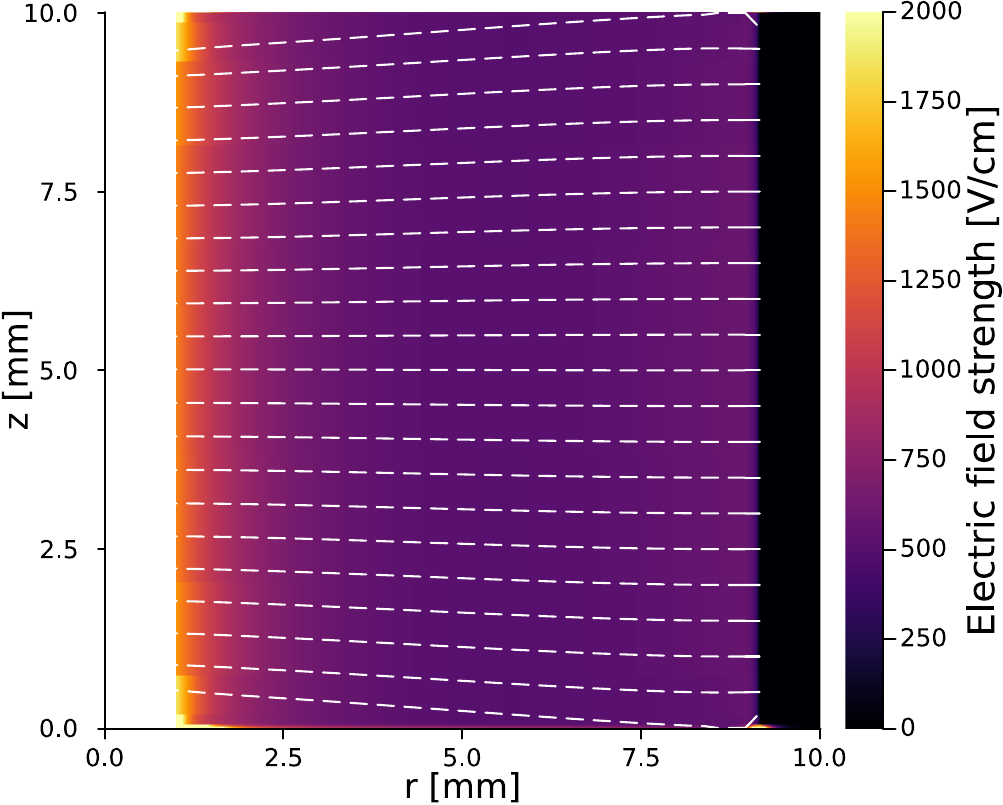}
    }
    \subfigure[]{
        \includegraphics[height=4.4cm]{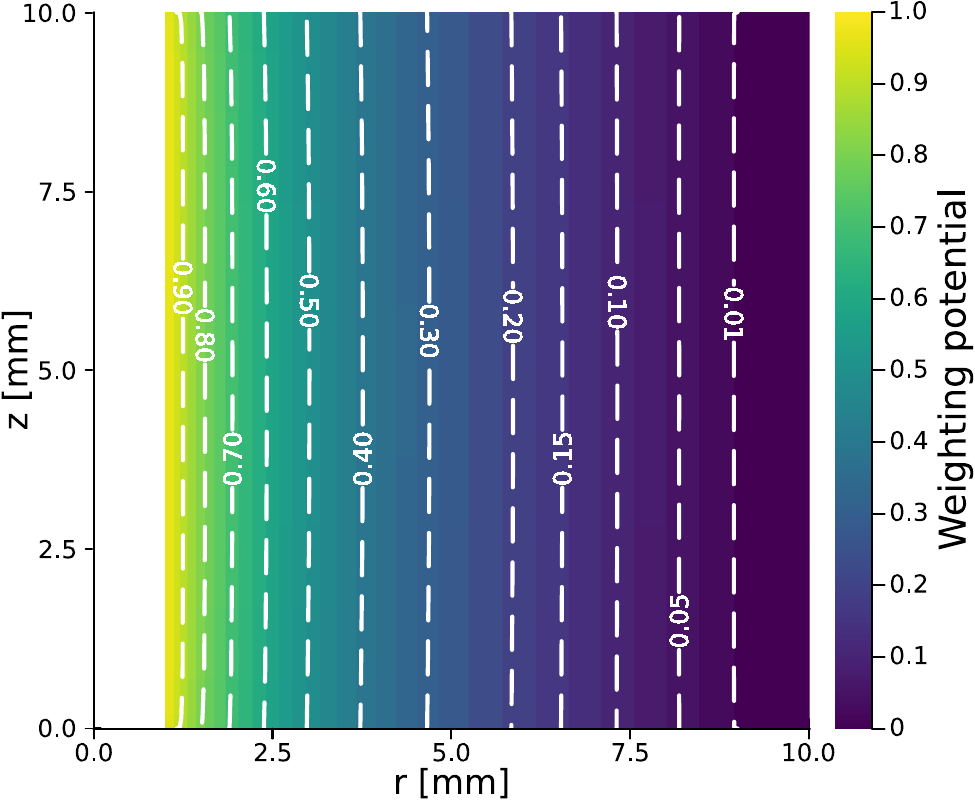}
    }
    \caption{
        The electric potential (dashed lines are the equipotential contours) (a),
        the electric field (white lines are the electric field lines) (b),
        and the weighting potential of the $p^+$ contact (dashed lines are the equipotential contours) (c)
        of the true-coaxial HPGe detector, as calculated with \emph{SSD.jl}
    }
    \label{fig.truecoaxial_fields}
\end{figure*}

The $n^+$ electrode of the detector is modeled with
lithium impurity doping via thermal diffusion.
We assume that the surface lithium concentration
is at its saturation level,
the annealing time $t_\text{an}$ is 18\,minutes,
the annealing temperature $T_\text{an}$ is 623\,K, 
the acceptor impurity density is 1$\times$10$^{10}$\,cm$^{-3}$,
and the bias voltage is 500\,V.
These parameters result in an $n$-type region
with a depth of \textasciitilde\,1\,mm.
Figure \ref{fig.truecoaxial_fields} shows the electric potential, weighting potential, and electric field, as calculated with \emph{SSD.jl}.
The center point in the z direction (z = 5\,mm)
exhibits sufficient uniformity in the z direction,
which allows us to use the one-dimensional analytical method
to calculate the CCE curve of the RCC layer
and compare it with the simulated CCE curve.
\subsection{Charge collection efficiency validation}\label{sec3.2}
We use both the numerical simulation and the analytical approach to
compute the drift of hole clouds
generated by events that all have a deposited energy of 10\,keV
(corresponding to 3389\,holes). All holes in the charge cloud of a single event start from the same position simultaneously for both two approaches.

\begin{figure}[!htb]
    \centering
    \includegraphics[width=1\hsize]{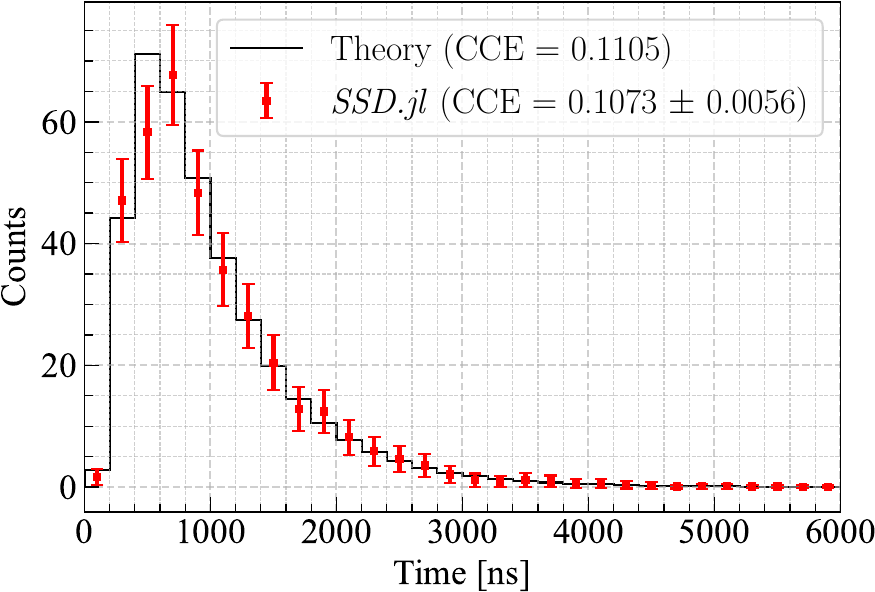}
    \caption{
         Analytically calculated (black) and numerically simulated (red) distributions of the arrival times of holes at the $p$-$n$ boundary. The holes originate from a single-site event with an energy of 10\,keV deposited at a depth of 0.5\,mm from the surface. The time bin width is set at 200\,ns to achieve adequate counting statistics for the simulated results, although the analytical results were calculated with a bin width of 1\,ns.
     }
    \label{fig.truecoaxial_time_pdf}
\end{figure}

Figure \ref{fig.truecoaxial_time_pdf} shows
the distribution of the time when the holes arrive at the $p$-$n$ boundary for a starting depth of 0.5\,mm from the surface. The integral of this curve divided by the initial hole number (3389)
is the charge collection efficiency of the event, see Equation \ref{eq.cce_calc}.
The simulated distribution agrees with the analytical one. Notably, the uncertainties in the simulated histogram are overestimated, similar to the calculation of CCE uncertainty described in Section \ref{sec2.4}.

\begin{figure}[!htb]
    \centering
    \includegraphics[width=1\hsize]{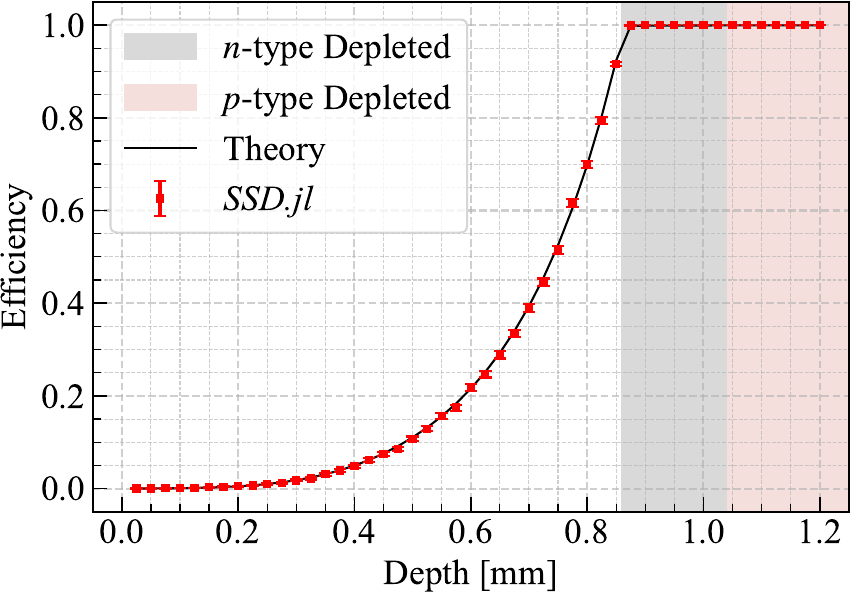}
    \caption{
         Analytically calculated (black) and numerically simulated (red) CCE curves of the RCC layer, calculated by generating events at different depths with an energy of 10\,keV.
     }
    \label{fig.truecoaxial_cce}
\end{figure}

In addition to the arrival time, we compare the depth dependency of the CCE by generating events at different depths in the RCC layer. Figure \ref{fig.truecoaxial_cce} shows that the two CCE curves from the numerical simulation and analytical solution are also consistent.

\subsection{Pulse shape validation}\label{sec3.3}

\begin{figure}[!htb]
    \centering
    \includegraphics[width=1\hsize]{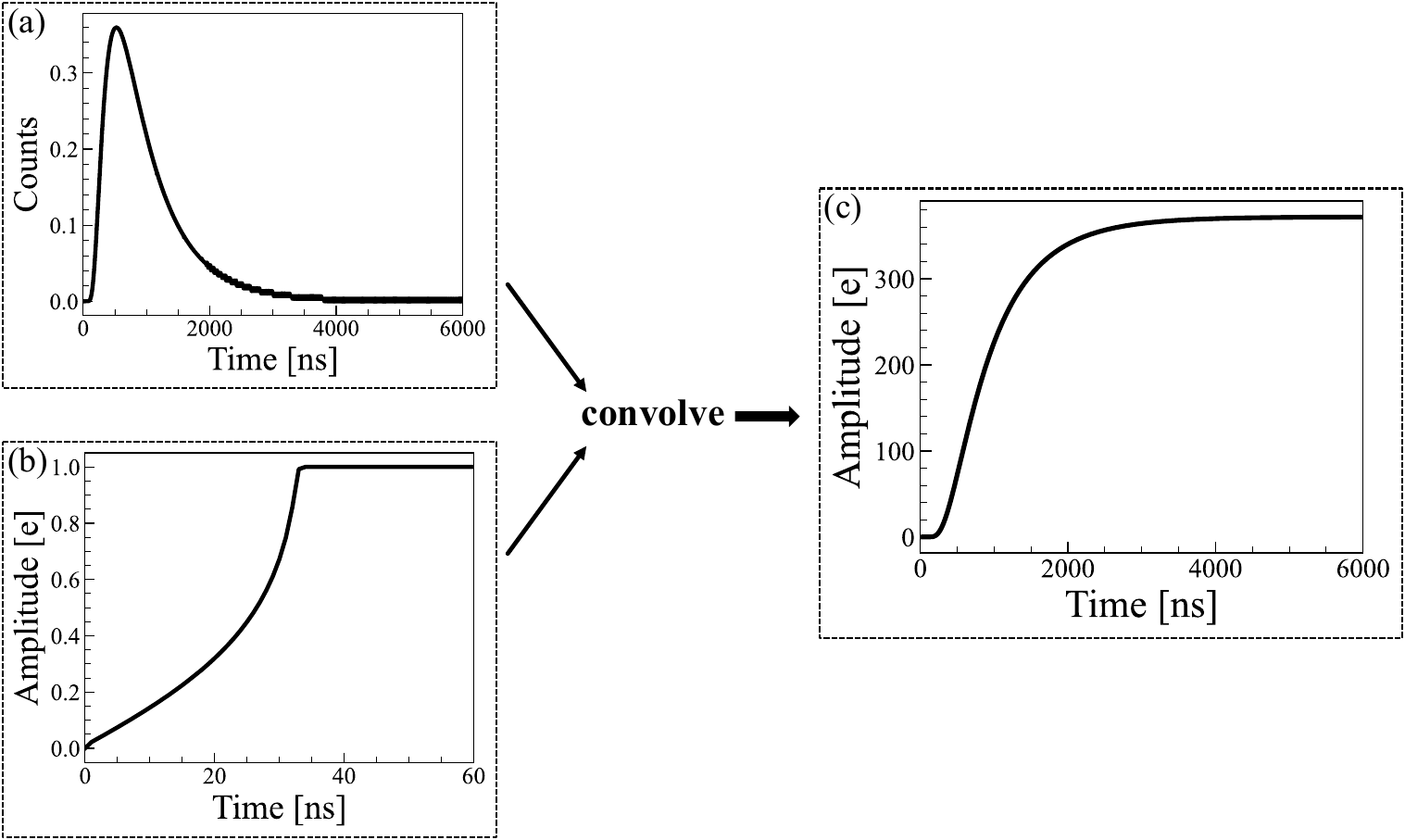}
    \caption{
        Illustration of the pulse shape calculation via the analytical method.
        (a) Distribution of hole arrival time at the $p$-$n$ boundary (the time bin width is 1\,ns).
        (b) The pulse shape of a single hole originating from the $p$-$n$ boundary.
        (c) The pulse shape of the event originating from the RCC layer.
    }
    \label{fig.truecoaxial_pulse_convolve}
\end{figure}

\begin{figure}[!htb]
    \centering
    \includegraphics[width=1\hsize]
    {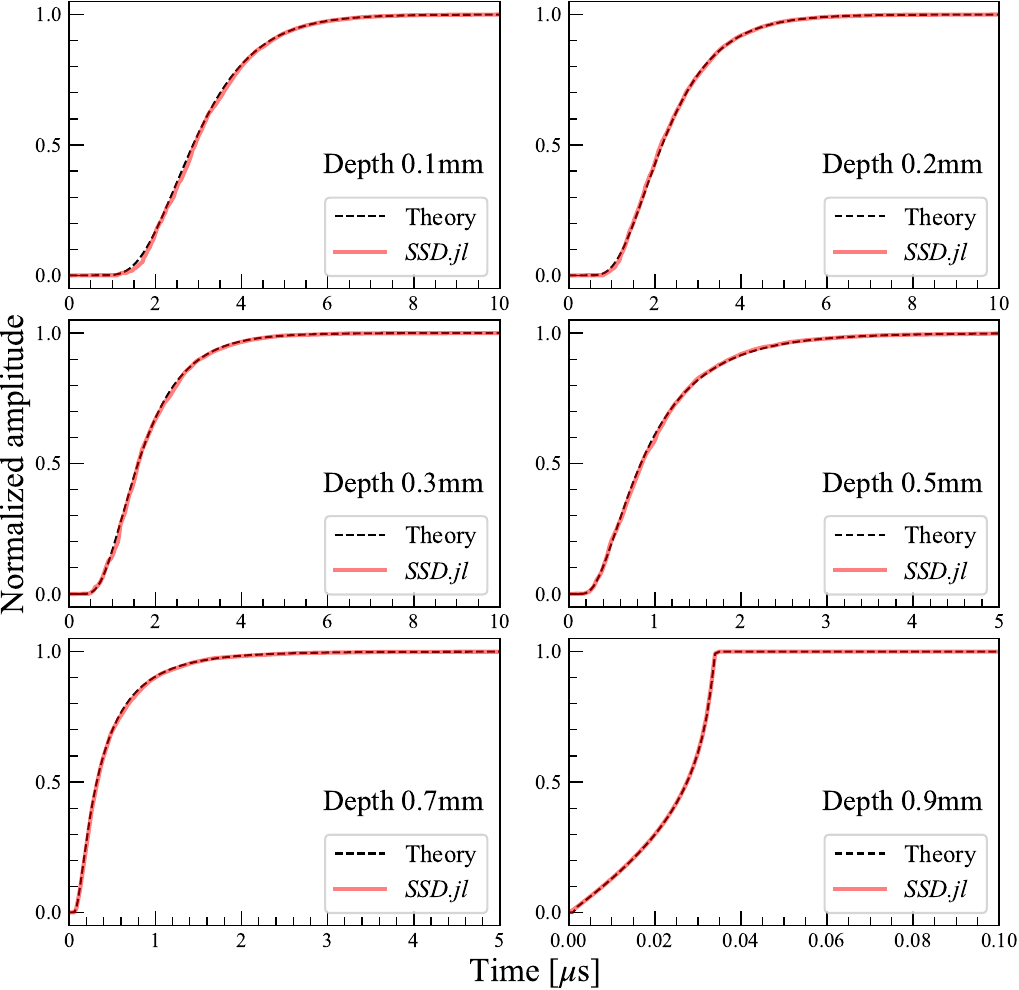}
    \caption{
        Analytically calculated (black) and numerically simulated (red) pulse shapes of 
        events originate at different depths
        (0.1, 0.2, 0.3, 0.5, 0.7, 0.9\,mm) in the RCC layer.
        Note the shorter time ranges for larger depths (closer to the sensitive region). 
    }
    \label{fig.truecoaxial_pulses}
\end{figure}

The numerical simulation generates pulse shapes directly. We can also generate semi-analytical pulse shapes by convolving the analytically derived distribution of hole arrival time at the $p$-$n$ boundary with a numerically simulated pulse of a single hole originating exactly at the $p$-$n$ boundary~\cite{bjorn_lehnert_search_2016}, as demonstrated in Figure \ref{fig.truecoaxial_pulse_convolve}.
Figure \ref{fig.truecoaxial_pulses} compares the numerical and semi-analytical pulse shapes for events generated at various depths in the RCC layer. Again, we see very good consistency between the simulation and analytical approaches.

\section{Results and discussions}\label{sec4}
\subsection{Experimental setup and detector fields}\label{sec4.1}
\begin{figure}[!htb]
    \centering
    \includegraphics[width=1\hsize]{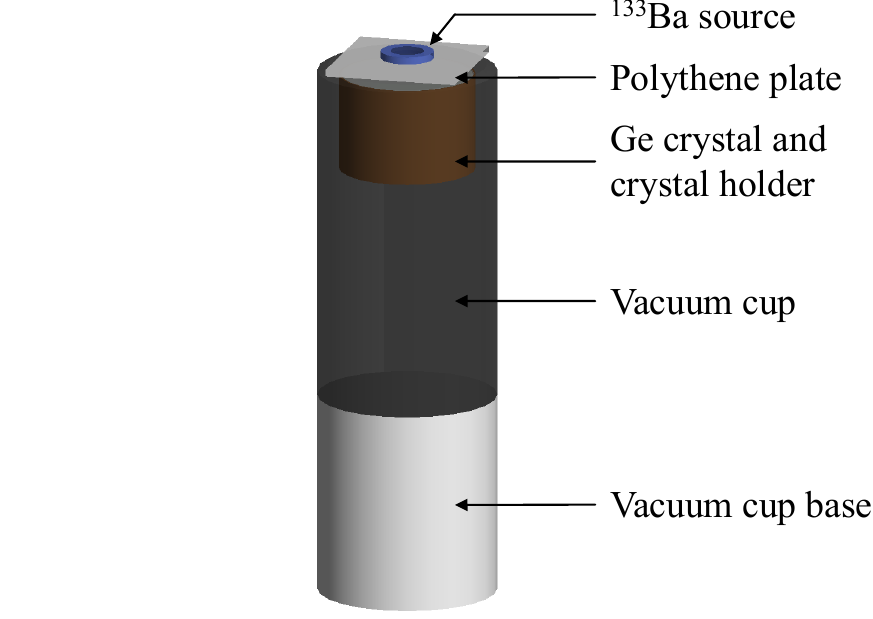}
    \caption{
        Experimental setup. The $^{133}$Ba source was centered at the top of the detector. A polythene plate was placed between the source and the detector to reduce pile-up events.
    }
    \label{fig.bege_det_geant4}
\end{figure}

We conducted a measurement on a
$p$-type Broad Energy Germanium detector (BEGe)
with an uncollimated $^{133}$Ba source centered at the top of the detector,
see Figure \ref{fig.bege_det_geant4}.
The $n^+$ electrode of the detector was heavily
doped with lithium impurities via thermal diffusion.
The detector parameters are listed in Table \ref{tab1}~\cite{dai_pulse_2024,dai_modeling_2023}.
The acceptor impurity density constant was derived by the experimental calibration.
The hole and electron lifetimes in the RCC layer are assumed to be identical in the experimental study.

\begin{figure}[htb]
    \centering
    \includegraphics[width=0.6\hsize]{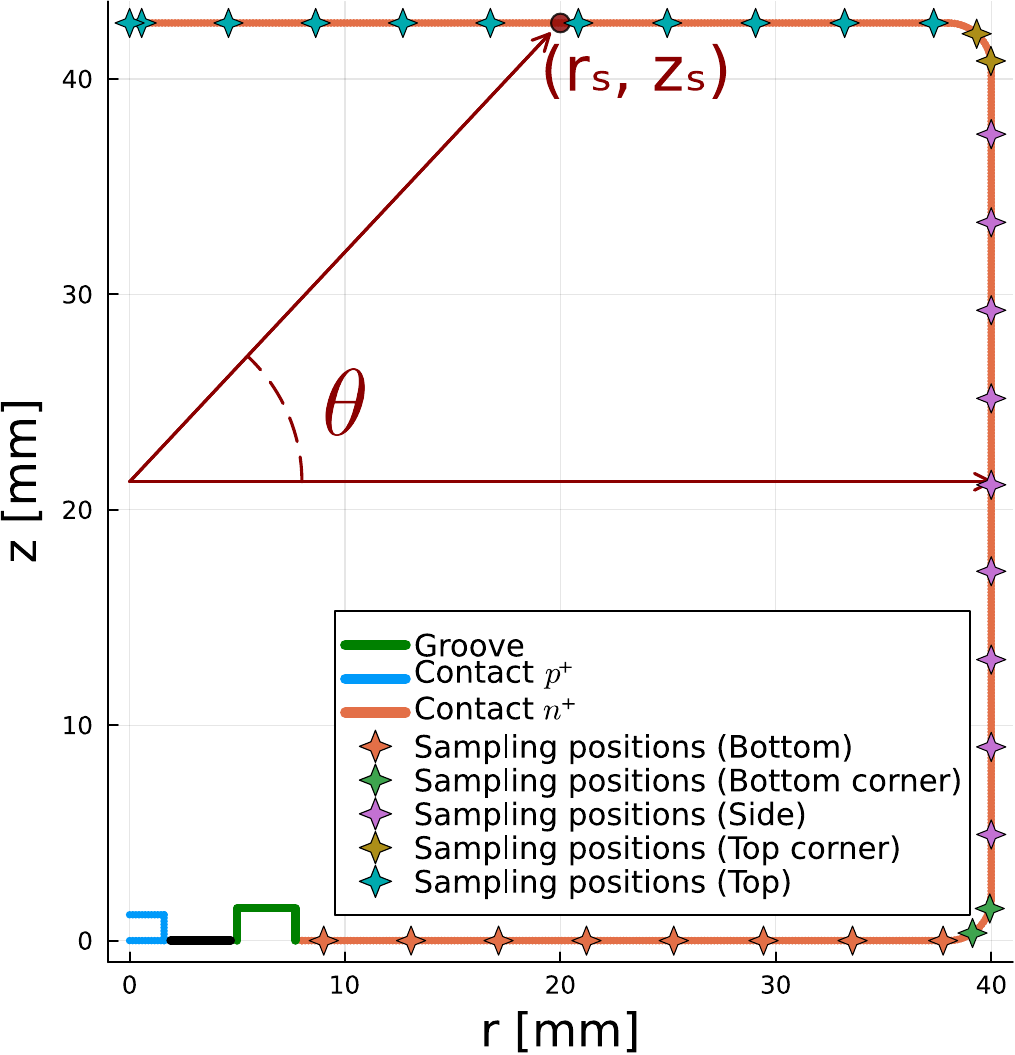}
    \caption{
        The r-z cross-section of the detector crystal. Also shown are the surface sampling positions (from the bottom, bottom corner, side, top corner, and top) that we use as charge cloud starting points in several following plots, and our definition of the center-to-surface angle $\theta$, see Equation \ref{eq.theta}.
    }
    \label{fig.bege_det_rz}
\end{figure}

Figure \ref{fig.bege_det_rz} shows the r-z cross-section of the crystal. It also shows the surface sampling points that we use in the following as starting points for several charge-drift simulations. The sampling positions include points
from the bottom, bottom corner, side,
top corner, and top,
which are used to display the results along the surface in several plots below.
The figure also shows the center-to-surface angle $\theta$ that we define as
\begin{equation}
\theta = \text{arctan}(\frac{\text{z}_\text{s}-\text{H}_\text{c}/2}{\text{r}_\text{s}})
\label{eq.theta}
\end{equation}
with radial and vertical coordinates $\text{r}_\text{s}$ and $\text{z}_\text{s}$ of the surface point,
and the crystal height $\text{H}_\text{c}$ ($=42.6$\,mm).

\begin{figure*}[!htb]
    \centering
    \subfigure[]{
        \includegraphics[height=4.7cm]{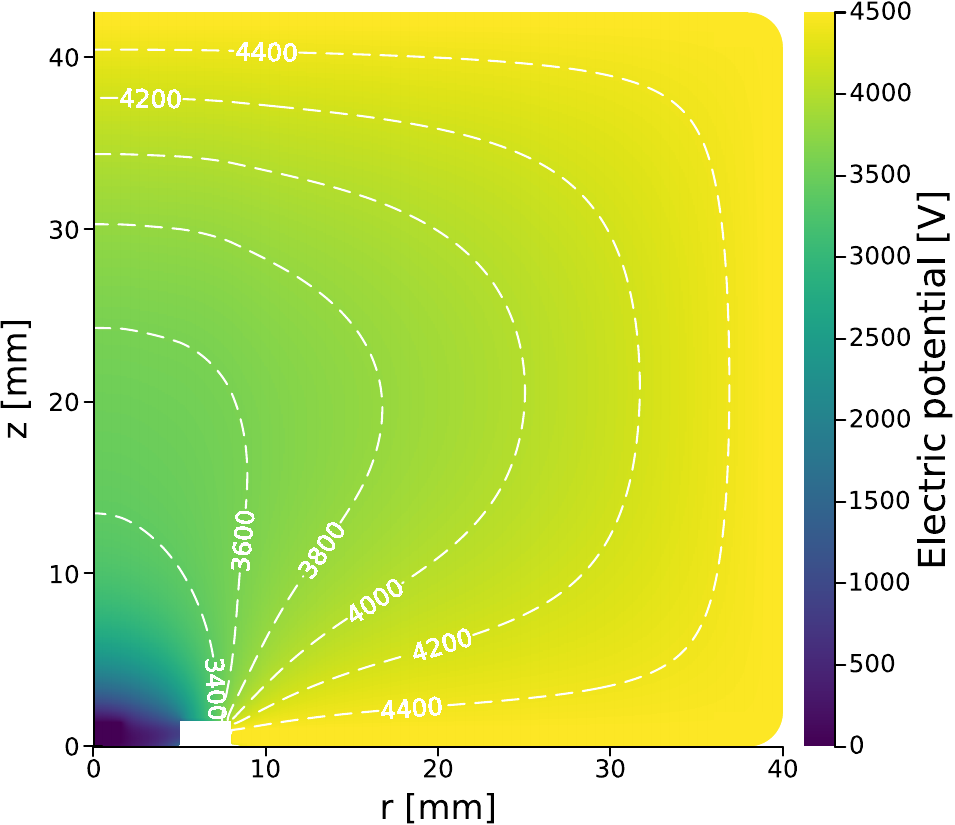}
    }
    \subfigure[]{
        \includegraphics[height=4.7cm]{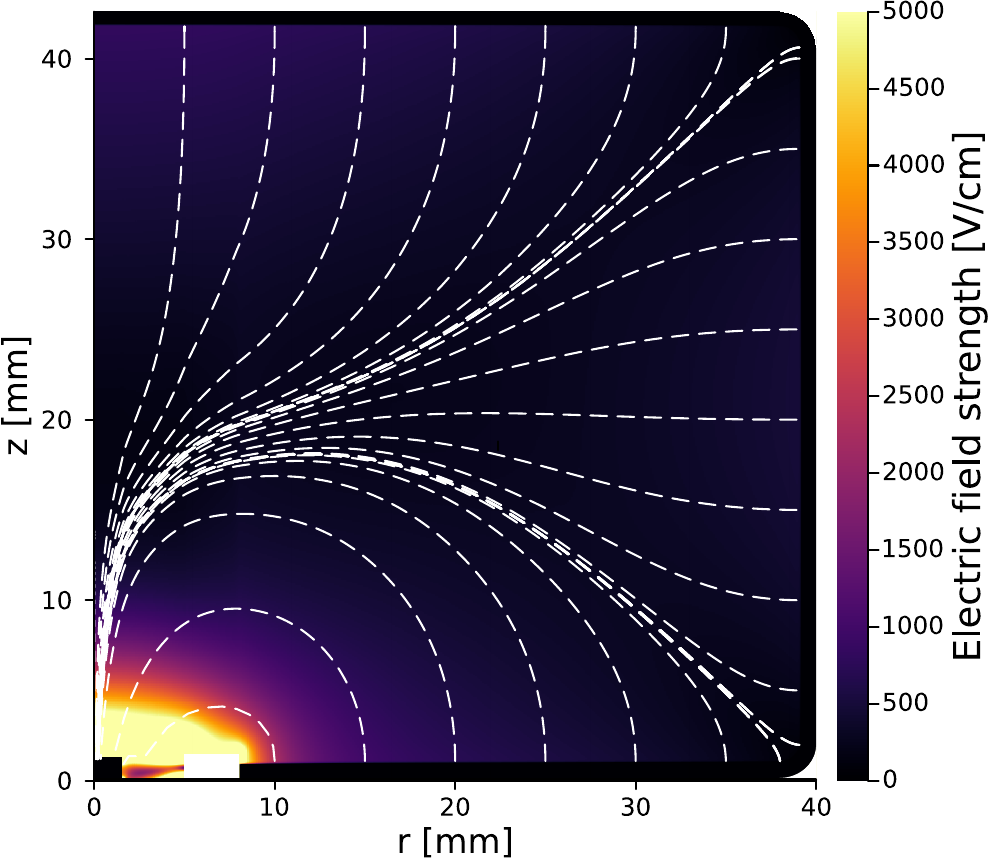}
    }
    \subfigure[]{
        \includegraphics[height=4.7cm]{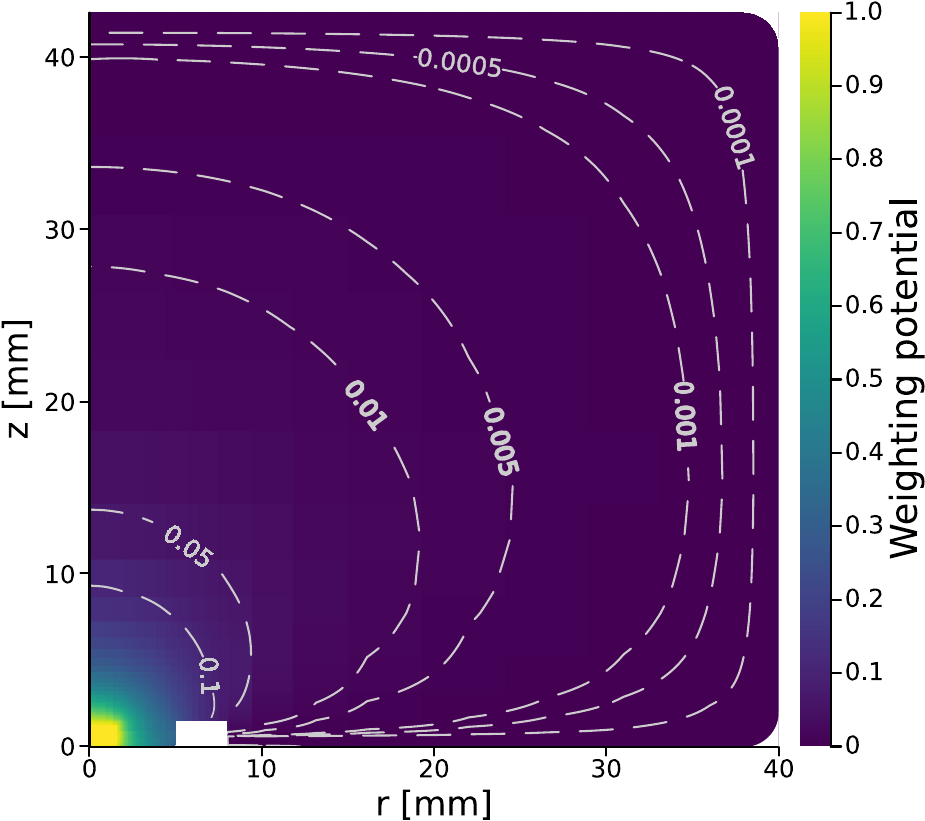}
    }
    \caption{
        The electric potential (dashed lines are the equipotential contours) (a),
        the electric field (white lines are the electric field lines) (b),
        and the weighting potential of the $p^+$ contact (dashed lines are the equipotential contours) (c)
        of the BEGe detector, as calculated with \emph{SSD.jl}.
    }
    \label{fig.bege_fields}
\end{figure*}

The electric potential, the electric field,
and the weighting potential of the BEGe detector are again calculated using \emph{SSD.jl},
see Figure \ref{fig.bege_fields}.
In the region near the $p^+$ contact, the electric field, the weighting potential,
and the gradient of the weighting potential are much larger than in the rest of the detector volume.

\begin{table}[t]
    \centering
    \caption{
    Parameters of the BEGe detector used in the simulation.
    }
    \label{tab1}
    \renewcommand\arraystretch{1.5}
    \scalebox{0.85}{
        \begin{tabular}{lll}
            \toprule
            Parameters & Values & Source\\
            \hline       
            Crystal radius & 40\,mm & Manufacturer \\
            Crystal height & 42.6\,mm & Manufacturer \\
            $p^+$ electrode hole radius & 1.6\,mm & Manufacturer \\
            $p^+$ electrode hole height & 1.2\,mm & Manufacturer \\
            Corner radius & 2\,mm & Manufacturer \\
            Groove inner radius & 5\,mm & Typical value \\
            Groove depth & 1.5\,mm & Typical value \\
            Groove outer radius & 8\,mm & Typical value \\
            Crystal temperature & 90\,K & Typical value \\
            Bias voltage & 4500\,V & Experiment \\
            Surface $n^+$ annealing temperature & 623\,K & Experiment \\
            Surface $n^+$ annealing time & 18\,min & Experiment \\
            Surface $n^+$ lithium density & 1.2$\times$10$^{17}$\,cm$^{-3}$ & Typical value \\
            Acceptor impurity density & 0.3$\times$10$^{10}$\,cm$^{- 3}$ & Experiment \\
            Sensitive region hole lifetime & 1\,ms & Typical value \\
            Sensitive region electron lifetime & 1\,ms & Typical value \\
            RCC layer hole lifetime & 800\,ns & Experiment \\
            RCC layer electron lifetime & 800\,ns & Experiment \\
          \bottomrule
        \end{tabular}
    }
\end{table}

\subsection{Drift paths and pulse shapes}\label{sec4.2}

\begin{figure}[!htb]
    \centering
    \subfigure[]{
        \includegraphics[width=0.7\hsize]{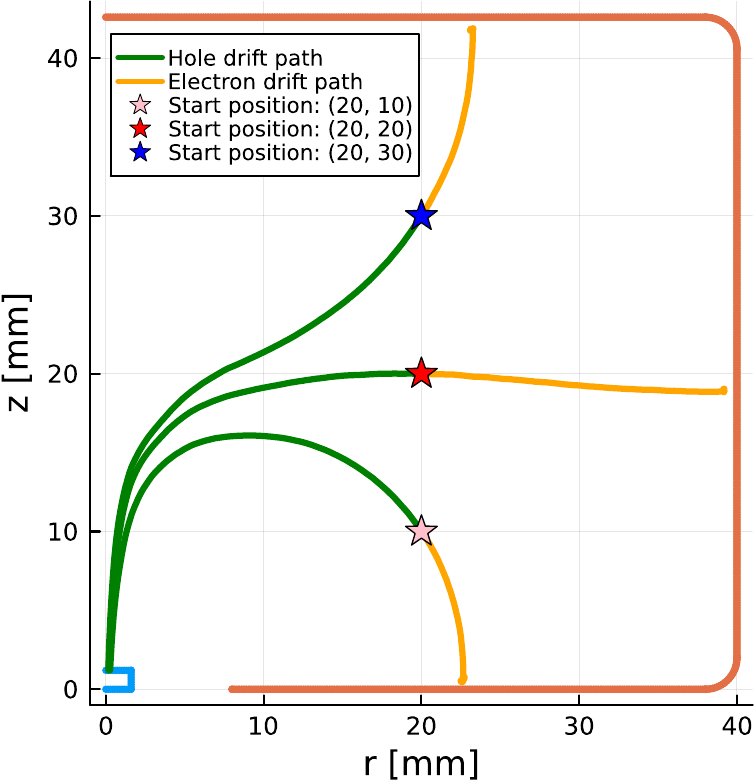}
    }
    \subfigure[]{
        \includegraphics[width=0.9\hsize]{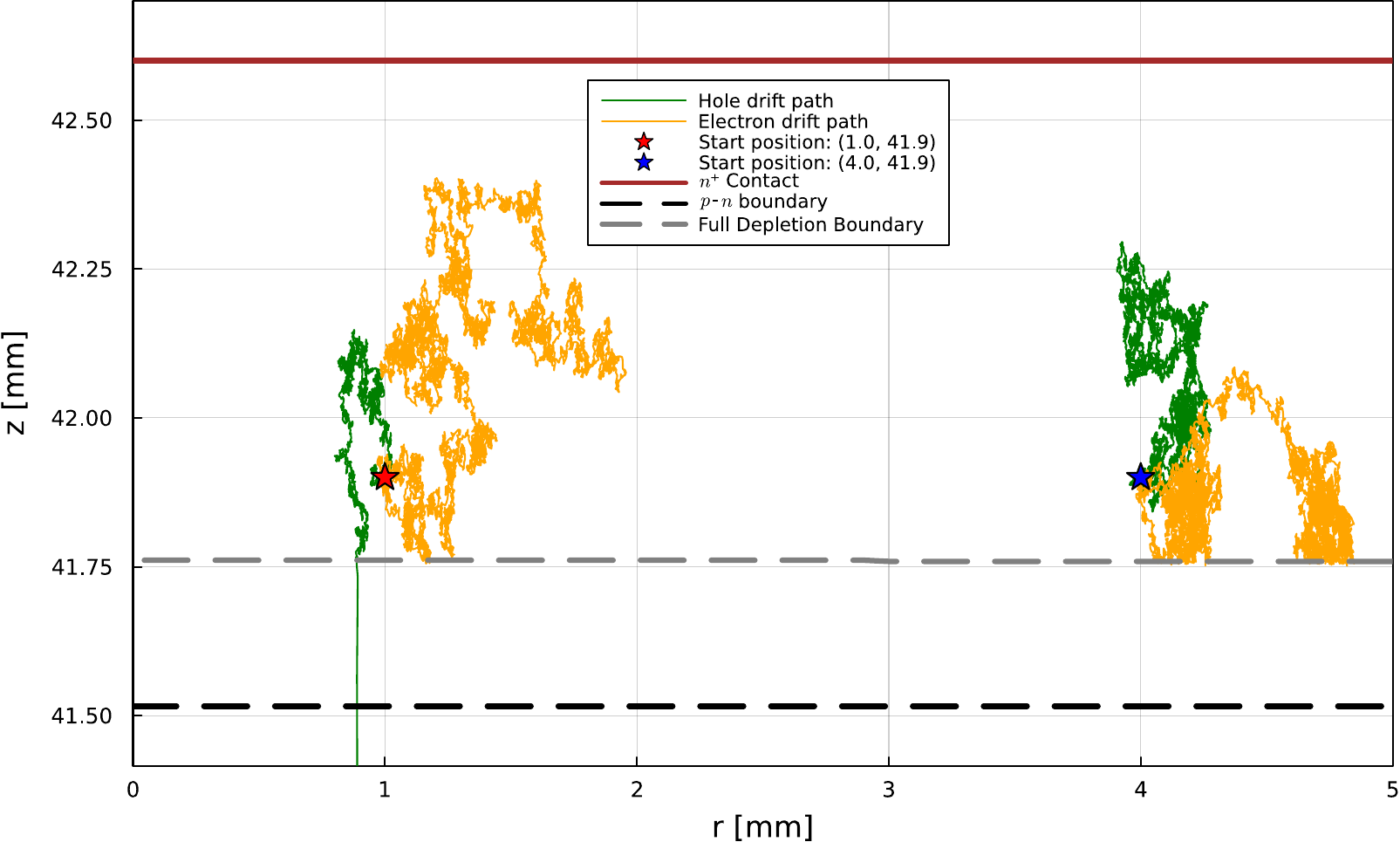}
    }
    \caption{
        Typical drift paths of three charge carrier pairs in the detector bulk (a)
        and typical drift paths of two charge carrier pairs in the RCC layer (b),
        simulated with \emph{SSD.jl}.
        For the case in the RCC layer, within a 10 $\upmu$s time window, the hole on the left reaches the depleted region and gets collected, while the one on the right is still diffusing.
    }
    \label{fig.bege_drift}
\end{figure}

Figure \ref{fig.bege_drift} illustrates the simulated charge carrier drift paths
in the sensitive region and in the RCC layer.
In the sensitive region, carriers approximately
drift along the electric field lines
before getting collected or drifting into the RCC layer.
In contrast, in the RCC layer,
carriers initially undergo a very slow random diffusion.
An electron will keep diffusing
until reaching the $n^+$ contact or getting trapped.
Even if it steps into the depleted region,
it will "bounce back" due to the electric field there.
However, for a hole,
once it steps into the depleted region,
it will drift rapidly along the electric field lines
until it is collected on the $p^+$ contact,
inducing a significant signal.

\begin{figure*}[!htb]
    \centering
    \includegraphics[width=0.83\hsize]{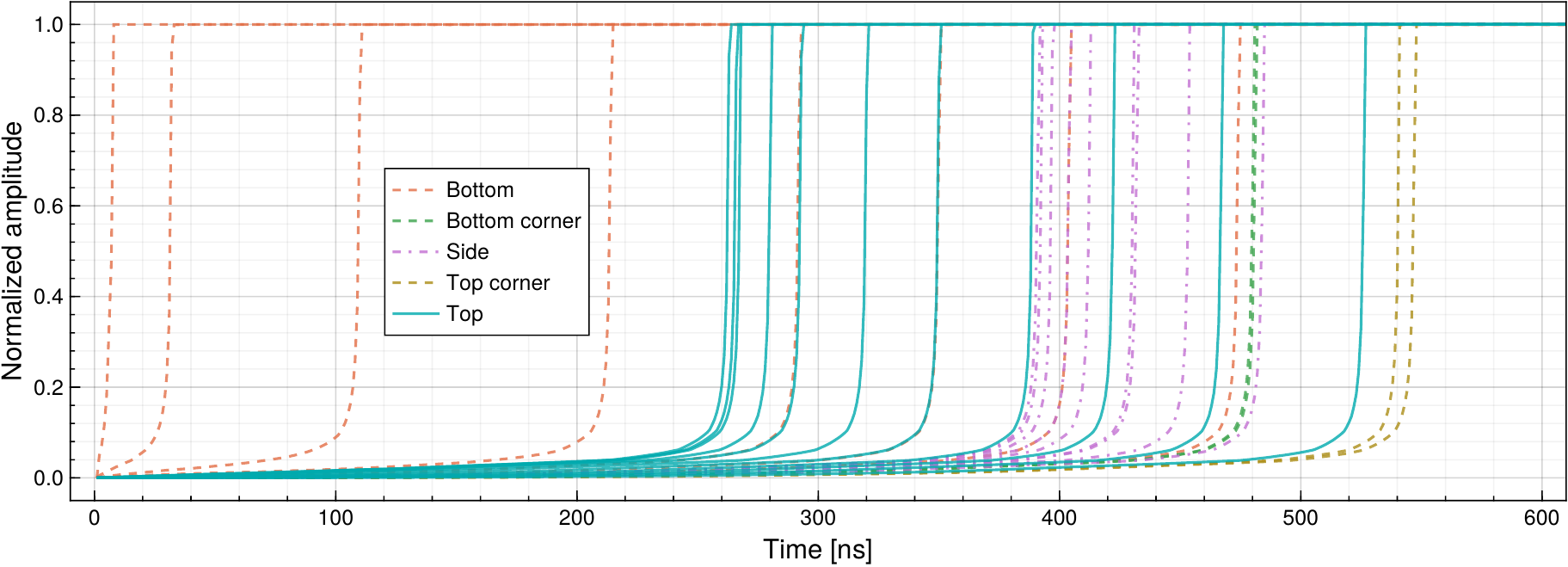}
    \caption{
        The simulated pulse shapes of events originated 
        at the $p$-$n$ boundary (in the sensitive region)
        at the sampling positions which are
        annotated in Figure \ref{fig.bege_det_rz}.
    }
    \label{fig.bege_bulk_pulses}
\end{figure*}

\begin{figure*}[!htb]
    \centering
    \includegraphics[width=0.83\hsize]{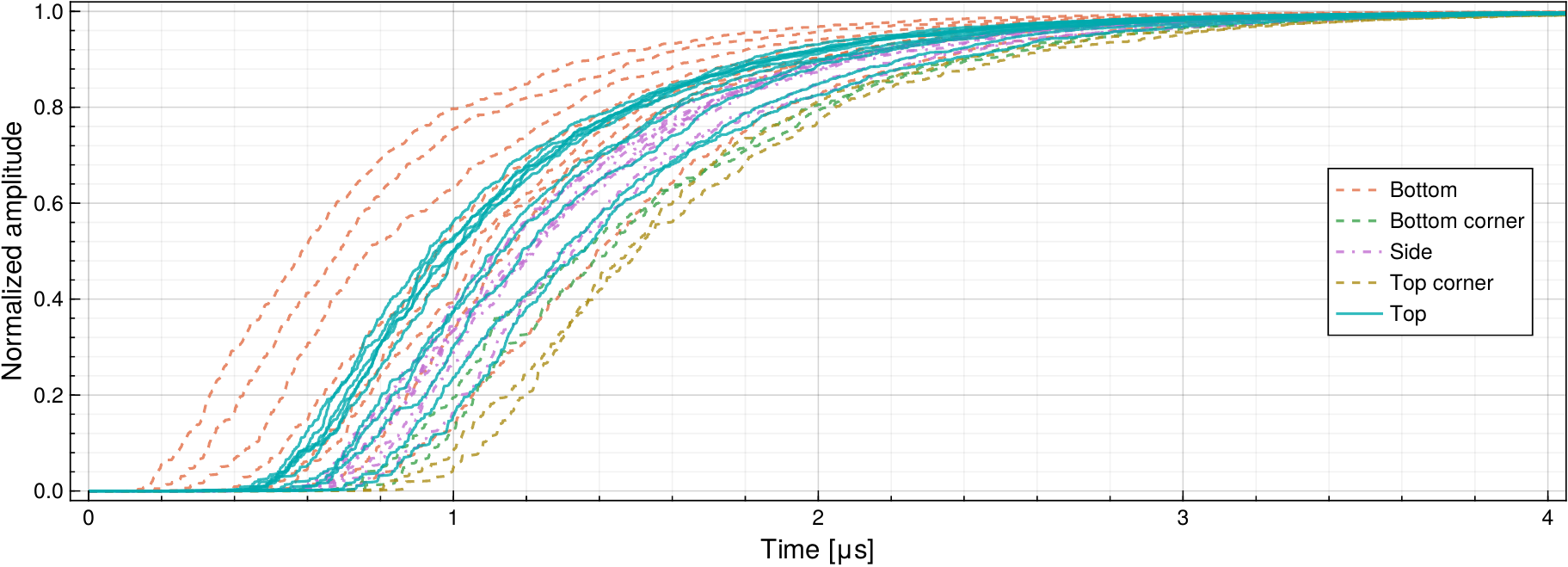}
    \caption{
        The simulated pulse shapes of events originated at a depth of 0.5\,mm from the surface
        (in the RCC layer)
        at the sampling positions which are 
        annotated in Figure \ref{fig.bege_det_rz}.
    }
    \label{fig.bege_surface_pulses}
\end{figure*}

Figures \ref{fig.bege_bulk_pulses} and
\ref{fig.bege_surface_pulses} depict
the normalized simulated pulse shapes of events originating
at the $p$-$n$ boundary (located in the sensitive region)
and at a depth of 0.5\,mm from the surface
(located in the RCC layer), respectively,
at the sampling positions annotated in
Figure \ref{fig.bege_det_rz}. 
Electronic response and electrical noise are not simulated.
These pulse shapes are divided into five batches
with regard to where the event originates:
the bottom, bottom corner, side, top corner, and top.

In most areas (far away from the $p^+$ contact),
the pulse shapes from the $p$-$n$ boundary
first undergo a slow rise,
followed by a rapid rise.
This rapid rise is caused by holes drifting
into the high-weighting-potential-gradient
and high-electric-field region,
which is near the $p^+$ contact in this case,
see Figure \ref{fig.bege_fields}.
The moment of the rapid rise
depends on the hit position of the event.
Generally, the rise will occur earlier
when the position is closer to the $p^+$ contact.

The position dependence is also evident
in the pulse shapes from the RCC layer,
despite the rise of the pulse shape being slowed down
by the random diffusion of carriers.
The fast rising edges of the pulse shapes
of the RCC layer events last about 3\,$\upmu$s,
while those from the $p$-$n$ boundary last only about 10\,ns.
For real detector pulse shapes,
which integrate the electronic response and electrical noise,
the slow rise may become invisible,
leaving only the fast-rising edge.
During the process of triggering and signal digitization,
we will also lose the information
of the rise start time of the pulse shape
due to the electrical noise.

\subsection{Non-uniformity of the RCC layer}\label{sec4.3}

\begin{figure}[!htb]
    \centering
    \includegraphics[width=1\hsize]{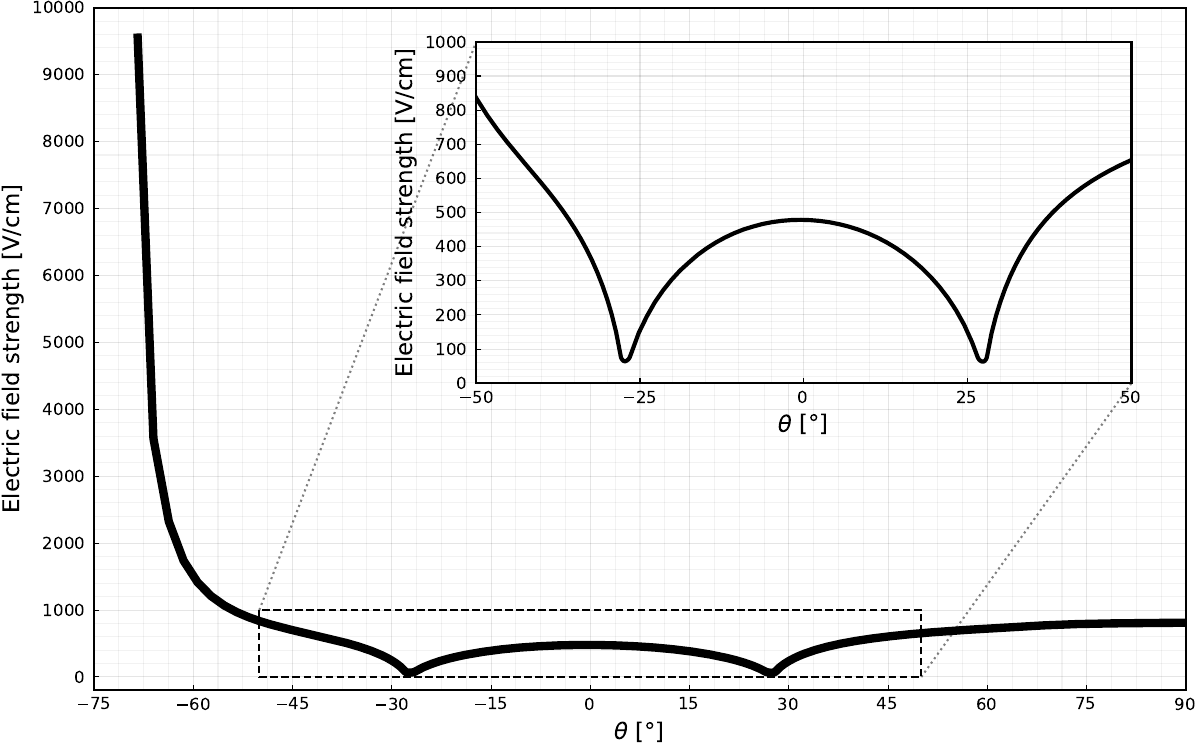}
   
    \caption{
        The simulated electric field strength on the $p$-$n$ boundary as a function of $\theta$. 
    }
    \label{fig.bege_pn_ef}
\end{figure}

The electric field strength along the $p$-$n$ boundary as a function of $\theta$,
from the groove to the top center,
is presented in Figure \ref{fig.bege_pn_ef}.
It varies within 200--800\,V/cm across most positions on the side
and top ($\theta \in [-45^{\circ},\ 90^{\circ}]$).
At the bottom part of the $p$-$n$ boundary, the electric field strength
increases steeply near the groove,
reaching \textasciitilde\,\SI{10000}\,V/cm at the groove.
The lowest electric field strength at the $p$-$n$ boundary (\textasciitilde\,60\,V/cm)
occurs at the detector corners ($\theta$ around $\pm 30^{\circ}$).
The $p$-$n$ boundary electric field strength at
the top center point is \textasciitilde\,800\,V/cm.

\begin{figure}[!htb]
    \centering
    \includegraphics[width=1\hsize]{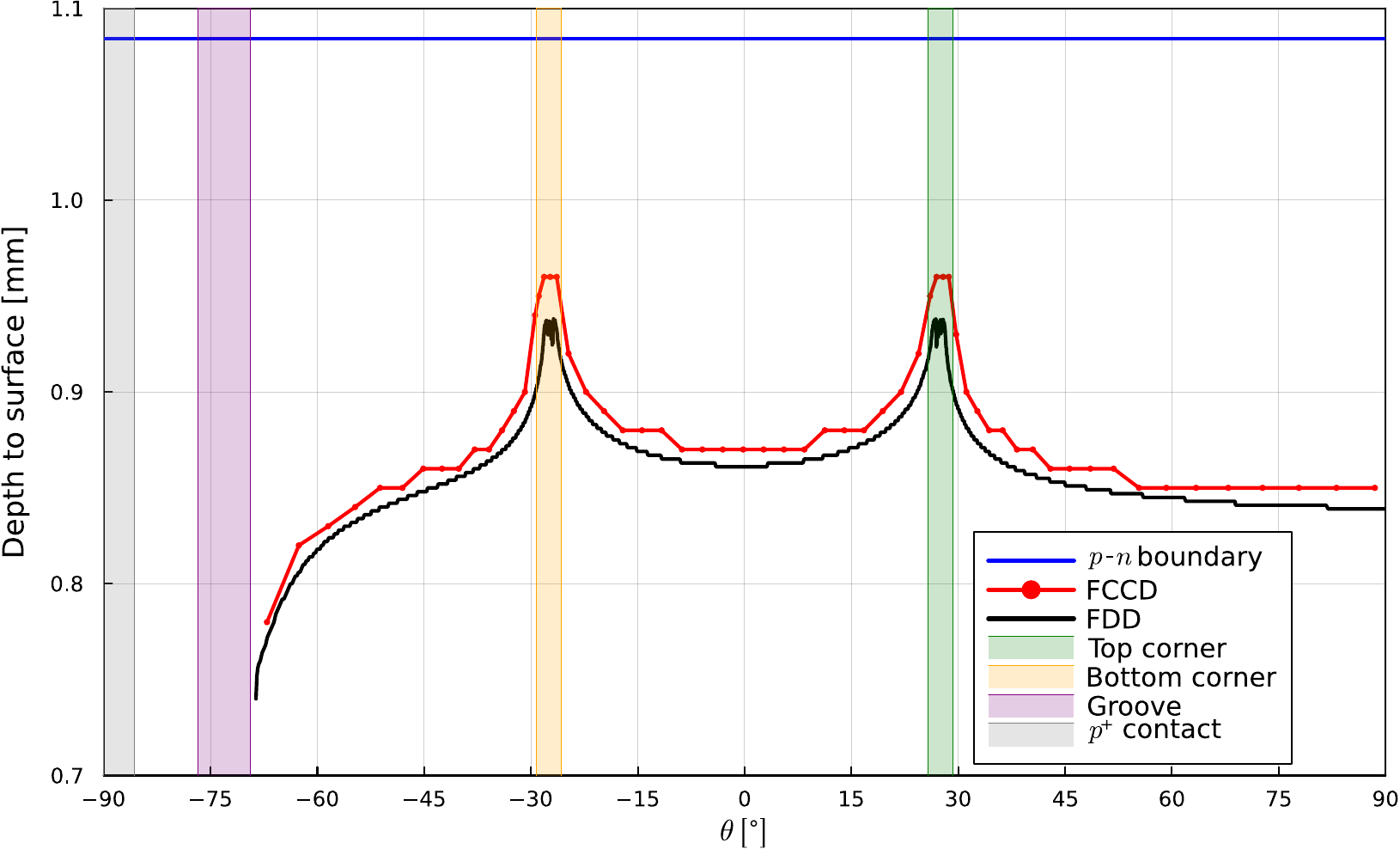}
    \caption{
    The simulated FDD (black) and FCCD (red), 
    and the $p$-$n$ boundary depth (blue) as a function of $\theta$.
    Surface areas of the BEGe detector are also shown: $p^+$ contact (gray), groove (purple), bottom corner (yellow), top corner (green), and the $p$-$n$ boundary (blue).
    }
    \label{fig.bege_pn_fccd_fdd}
\end{figure}

Figure \ref{fig.bege_pn_fccd_fdd} presents the simulated FDD, FCCD
and the depth of the $p$-$n$ boundary against $\theta$.
The predicted full-depletion boundary is
slightly shallower than the full CCE boundary,
indicating that incomplete charge collection also
occurs in the depleted region near the full-depletion boundary,
due to the weak electric field and random diffusion of carriers.
Figure \ref{fig.bege_pn_ef} and \ref{fig.bege_pn_fccd_fdd} demonstrate
a significant correlation between the electric field
at the $p$-$n$ boundary and the properties of the RCC layer,
namely, a stronger electric field corresponds to
a larger depleted region and a larger full CCE region.
At the top center ($\theta = 90^{\circ}$),
the simulated FDD and FCCD are determined to be
(839$\pm$2)\,$\upmu$m and (850$\pm$10)\,$\upmu$m, respectively.
The latter is consistent with the experimentally measured value of
(870$\pm$67)\,$\upmu$m~\cite{dai_modeling_2023}.
It should be noted that the uncertainties reported for the simulated FDD and FCCD reflect only
the numerical precision or statistical uncertainty and do not account for systematic uncertainties
arising from limited knowledege of detector parameters.
These systematic uncertainties may be dominant and significantly larger than
the values presented here.

\subsection{CCE-corrected simulated spectrum}\label{sec4.4}

\begin{figure*}[!htb]
    \centering
    \subfigure[]{
        \includegraphics[width=0.32\hsize]{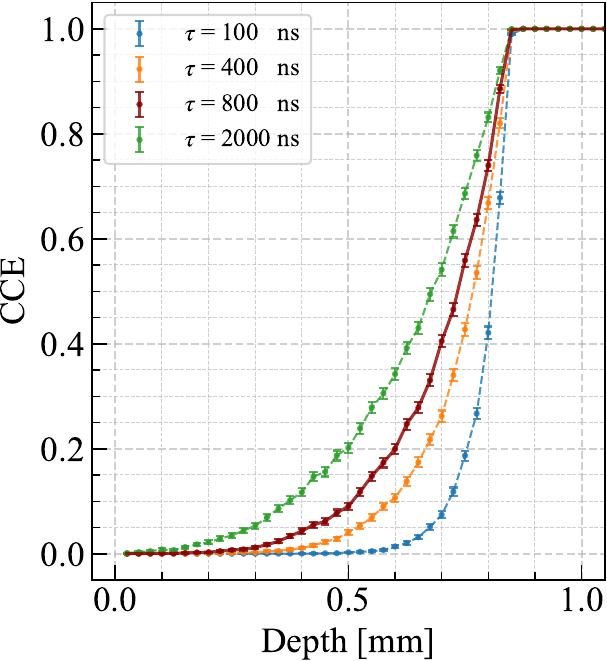}
    }
    \subfigure[]{
        \includegraphics[width=0.32\hsize]{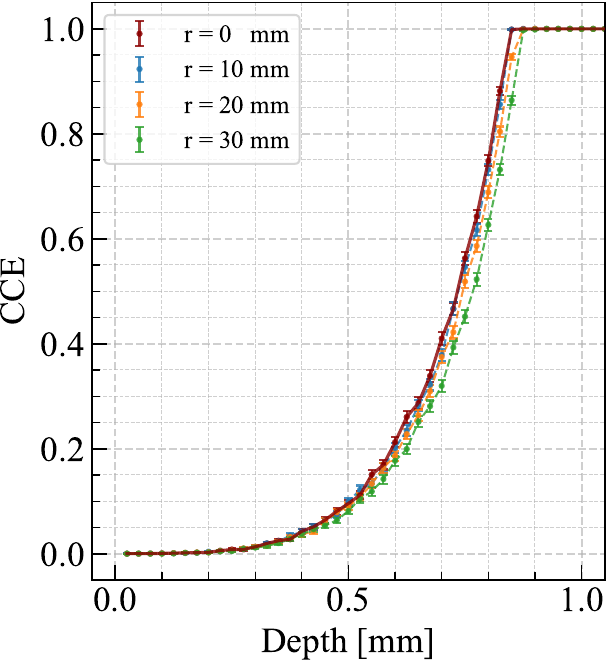}
    }
    \subfigure[]{
        \includegraphics[width=0.32\hsize]{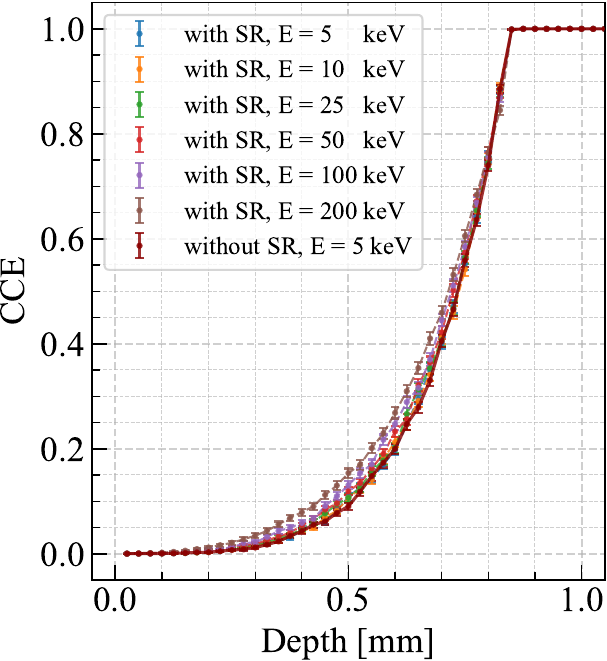}
    }
    \caption{
        (a) CCE curves for events at the top center point of the detector (r = 0\,mm)
        with varying RCC layer carrier lifetimes ($\tau$),
        a simulated energy (E) of 5\,keV,
        and no self-repulsion (SR).
        (b) CCE curves for events at different detector positions around the top center,
        with $\tau$ = 800\,ns, no SR, and E = 5\,keV. 
        (c) CCE curves with and without the self-repulsion,
        while $\tau$ = 800\,ns and r = 0\,mm.
        }
    \label{fig.bege_cces}
\end{figure*}

Figure \ref{fig.bege_cces}a displays the simulated CCE curves for events at the top center of the detector
with varying RCC layer carrier lifetimes ($\tau$), a simulated energy (E) of 5\,keV, and no self-repulsion effect.
Figure \ref{fig.bege_cces}b presents the simulated CCE curves
for events at different detector positions around the top center
with $\tau$ = 800\,ns, E = 5\,keV, and no self-repulsion effect.

When the self-repulsion effect is not simulated, the carriers drift independently.
Therefore, the simulated event energy affects the estimated statistical uncertainty of the CCE but not its mean value, according to Equations \ref{eq.cce_calc}-\ref{eq.cce_err_calc}.
However, when the effect is simulated, it can influence the charge drift, thereby affecting the mean value of CCE.
Figure \ref{fig.bege_cces}c presents the simulated CCE curves with and without the self-repulsion effect considered,
while $\tau$ = 800\,ns, r = 0\,mm, and E = 5\,keV.
For the case with the self-repulsion effect, we also simulated the CCE curves with larger event energies in 0--200\,keV, also shown in Figure \ref{fig.bege_cces}c. The energy range was chosen to cover more than 99\% of the energies deposited in the RCC layer in this case.

These results demonstrate that in this case, the r-position variation and the self-repulsion effect have a minimal impact on the CCE curve compared to the RCC layer carrier lifetimes.
Therefore, to accelerate the computation, we ignored the self-repulsion effect
and used the CCE curve of the top center to represent the entire top RCC layer.

\begin{figure*}[!htb]
    \centering
    \includegraphics[width=1\hsize]{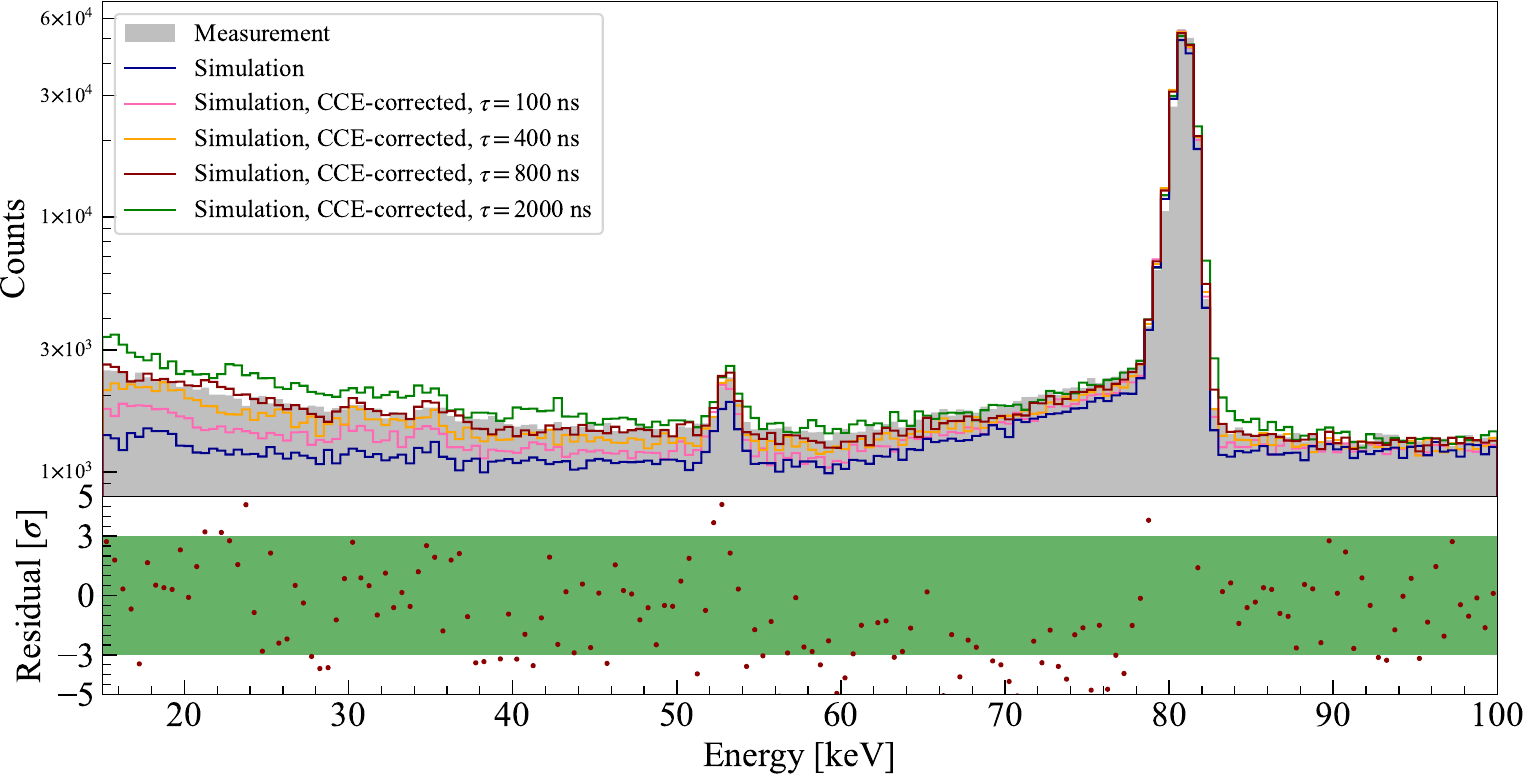}
    \caption{
        Experimentally measured energy spectrum (gray area), compared to
        simulated spectra that are corrected using the CCE determined
        with different RCC layer carrier lifetimes:
        800\,ns (solid red), 100\,ns (dashed pink),
        400\,ns (dashed orange) and 2000\,ns (dashed green),
        and simulated spectrum without CCE correction (solid blue).
        An RCC layer carrier lifetime of 800\,ns shows the best match to the measured spectrum, determined by $\chi^2$ minimization.
        The peaks around 31/35\,keV and 53/81\,keV originate from the X-rays (30.63/30.97/35.05/35.90\,keV)
        and $\gamma$-rays (53.16/79.61/81.00 keV) of the $^{133}$Ba source.
        The bottom half of the figure shows the residuals of the simulated spectrum for this 800\,ns lifetime to the measured spectrum.
    }
    \label{fig.bege_spectra_match}
\end{figure*}

By simulating energy depositions in the crystal with
Geant4 and applying corrections using the simulated CCE curve,
we obtained the CCE-corrected simulated energy spectrum.
The simulation-related parameters are also shown in Table \ref{tab1}.
The RCC layer carrier lifetimes
were determined to be 800\,ns
by matching the CCE-corrected simulated spectrum with
the experimental spectrum with $\chi^2$ minimization method.
As shown in Figure \ref{fig.bege_spectra_match},
the CCE-corrected simulated spectrum using $\tau_\text{e/h,I}$ = 800\,ns
agrees well with the experimental spectrum.
About 90\% of the simulated results lie within 5\,$\sigma$ of the experimental results (using the estimated Poisson fluctuation).

Minor discrepancies observed for some data points (mainly \textasciitilde\,70 keV) may arise from the oversimplification of the currently employed constant-lifetime trapping model, insufficient knowledge of detector parameters, or the neglect of the position dependence of CCE and self-repulsion effect, listed in descending order of likelihood.

Notably, the actual carrier lifetime is expected to vary with depth in the RCC layer, driven by a non-uniform distribution of trapping centers. This variation may originate from the crystal growth process or, more plausibly, from external contaminants introduced during fabrication. The density profile of such contaminants would likely exhibit a depth gradient, analogous to the lithium density variation shown in Figure \ref{fig.impurity_curves}. Implementing a depth-dependent-lifetime trapping model may significantly improve the simulation accuracy.

Therefore, the fitted lifetime reported here should be interpreted as an effective value subject to systematic uncertainties related to these factors. A comprehensive study addressing all these effects, including the implementation of the depth-dependent-lifetime trapping model and the variations in detector parameters, would require a massive number of simulations.

Due to the high computational cost associated with the fine grid resolution and carrier statistics in the current framework, such a study is impractical with the current framework. To address this performance bottleneck, we are currently developing a GPU-accelerated code. An extensive investigation utilizing this accelerated code is planned for future work.

\section{Summary and outlook}\label{sec5}
We have developed and verified a novel mechanistic three-dimensional pulse shape simulation method for events originating in the reduced charge collection (RCC) surface layer of $p$-type HPGe detectors.
This method is now publicly available as part of the open-source simulation package \emph{SolidStateDetectors.jl}.

In our approach, we model the ionized impurity density profile of the RCC layer based on the thermal diffusion process of lithium.
The hole and electron mobility models are based on a combination of ionized impurity scattering, neutral impurity scattering, and acoustic phonon scattering.
We use a random walk algorithm to simulate carrier diffusion and derive the diffusion coefficient from the mobility via the Einstein relation.
Optionally, we superimpose the charge-carrier self-repulsion effect.
Finally, we use a constant-lifetime trapping model to simulate carrier trapping in the RCC layer.

We have validated the simulation method against analytical calculations for a hypothetical true-coaxial HPGe detector. 
The time distributions for the arrival of holes at the $p$-$n$ boundary, the depth-dependency charge collection efficiency (CCE), and the pulse shapes all agree very well between both approaches.

We have also validated our method against a spectrum measured with an actual $p$-type BEGe detector.
The simulated RCC layer depth at top center of the detector is consistent with the experimental measurement.
We match the measured spectrum with a Monte-Carlo spectrum, with only minor discrepancies, when we simulate the charge collection using an RCC layer carrier lifetime of 800\,ns.
To achieve an even better agreement with the measured data, we need more accurate detector parameters, including geometry, lithium diffusion process parameters, and the acceptor impurity density profile.

So far, we have used a constant-lifetime trapping model.
The actual lifetime likely varies
with depth in the RCC layer due to the varying impurity density.
A depth-dependent-lifetime trapping model may improve the simulation results even further and will be a subject of future work.

The novel simulation method we presented here can serve as a basis for the investigation of RCC layer physics.
It also provides a new basis to develop and test novel pulse shape discrimination techniques, by generating realistic surface pulse libraries accompanied by ground truth data. This can help current and upcoming low-background experiments to use tighter background cuts with simulation-based control of systematics and cut efficiencies.

Although this paper focuses on $p$-type HPGe detectors, the method and code are also applicable to $p$-type multi-contact HPGe detectors.
Additionally, this framework provides a solid foundation for future extensions to $n$-type HPGe and other semiconductors (e.g., Silicon, CdZnTe) through adjustments to the doping profile and charge drift model.

\section*{Acknowledgments}
This work was supported by
the National Natural Science Foundation of China
(Grant No. 12425507 and No. 12175112),
the National Key Research and Development Program of China
(Grant No. 2023YFA1607101 and No. 2022YFA1604701),
and the PhD student fund for short-term overseas visits of Tsinghua University
(Contract Number: 2024088).

\bibliographystyle{spphys_three_authors.bst}
\bibliography{ref.bib}

\end{document}